\documentclass[aps,pra,twocolumn]{revtex4-1}

\usepackage[english]{babel}
\usepackage[T1]{fontenc}
\usepackage[utf8]{inputenc}

\usepackage{bbm}  
\usepackage{graphicx}
\usepackage{amsmath}
\usepackage{amssymb}
\usepackage{braket}
\usepackage{booktabs}
\usepackage{multirow}
\usepackage{simplewick}
\usepackage{bm} 
\usepackage{xcolor}
\usepackage{ulem}
\usepackage{threeparttable}

\usepackage{cellspace}
\newcommand\cincludegraphics[2][]{\raisebox{-0.4\height}{\includegraphics[#1]{#2}}}

\usepackage[autostyle=true]{csquotes}

\begin{document}                                                                                                                                                                   
                                                                                                                                                                                   
\title{Generation of Spin-Adapted and Spin-Complete Substitution Operators for\\
(High Spin) Open-Shell Coupled Cluster of Arbitrary Order}
\author{Nils Herrmann}
\email{N.Herrmann@uni-koeln.de}
\author{Michael Hanrath}
\email{Michael.Hanrath@uni-koeln.de}

\affiliation{%
Institute for Theoretical Chemistry, University of Cologne, 
Greinstra\ss e 4, 50939 Cologne, Germany
}%

\date{\today}

\begin{abstract}
A rigorous generation of spin-adapted (spin-free) substitution operators for high spin ($S=S_z$) references of arbitrary substitution order and spin quantum number $S$ is presented. The generated operators lead to linearly independent but non-orthogonal CSFs when applied to the reference and span the complete spin space. To incorporate spin completeness, spectating substitutions (as e.g. $\hat{E}_{iv}^{va}$) are introduced. The presented procedure utilizes L\"owdin's projection operator method of spin eigenfunction generation to ensure spin completeness. The generated operators are explicitly checked for (i) their linear independence and (ii) their spin completeness for up to 10-fold substitutions and up to a multiplicity of $2S+1 = 11$. A proof of concept implementation utilizing the generated operators in a coupled cluster (CC) calculation was successfully applied to the high spin states of the Boron atom. The results show pure spin states as well as small effects on the correlation energy compared to spinorbital CC. A comparison to spin-adapted but spin-incomplete CC shows a significant spin incompleteness error. 
\end{abstract}

\maketitle

\def\Op{\hat}
\def\<{\langle}
\def\>{\rangle}
\def\ls{\,\mathop{\text{ls}}\,}
\section{Introduction}
\label{Introduction}

The generalization of the coupled cluster (CC) \cite{Coester1958, Coester1958a} approach to high spin open-shell cases is an ongoing research field, where a lot of progress has been made in the last decades (see e.g. \cite{Mukherjee1975, Moitra1977, Nakatsuji1978, Nakatsuji1978a, Lindgren1978, Knowles1993, Knowles2000, Szalay1997, Heckert2006, Wilke2011, Nooijen1996, Janssen1991, Neogrady1992, Neogrady1994, Neogrady1995, Li1993, Li1994, Li1995, Li1995a, Li1995b, Jeziorski1995, Jankowski1999, Nooijen2001, Sen2012, Datta2008, Datta2009, Datta2013, Datta2019}). 

\begin{center}
\begin{figure}
\includegraphics[scale=1]{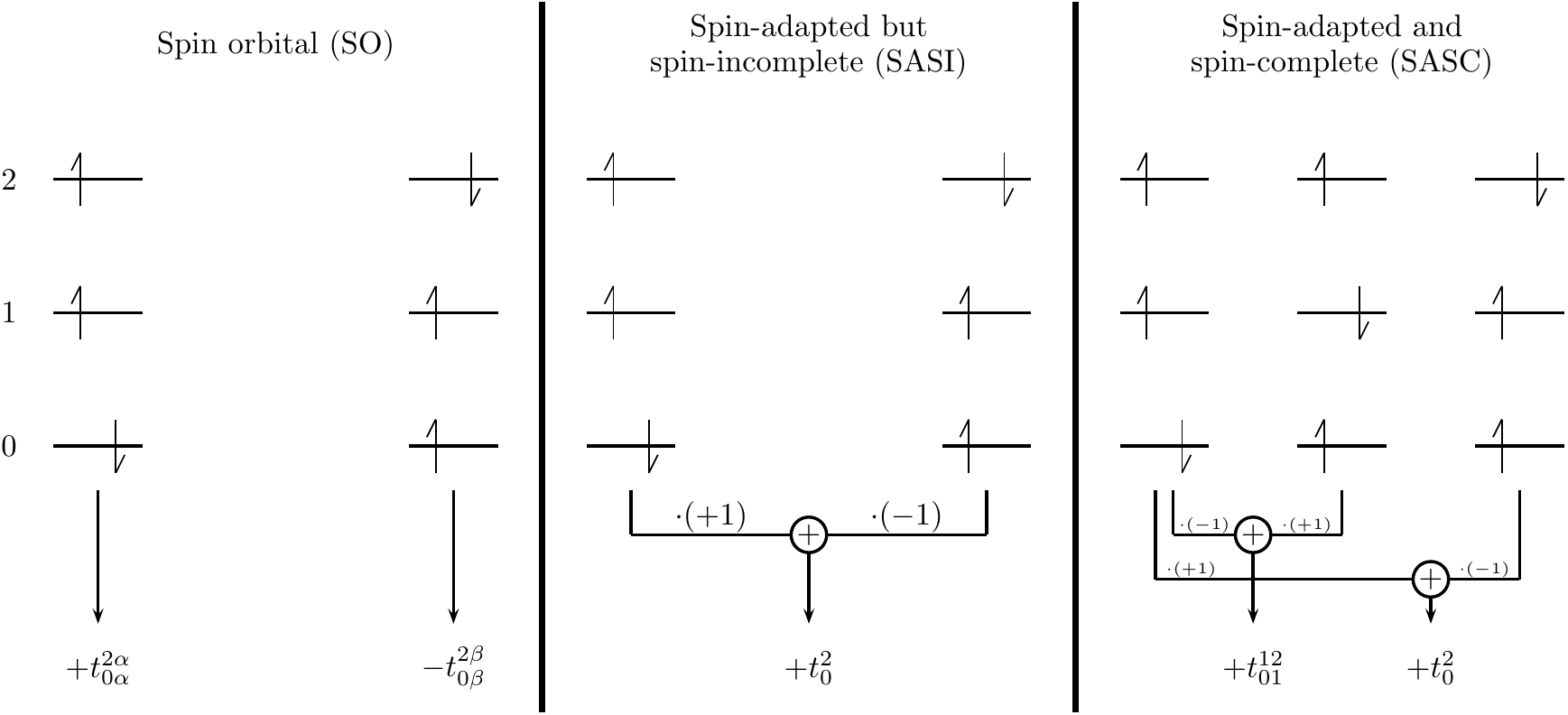}
\caption{\label{SOSISC.fig}Determinants and CSFs for the spatial configuration $\ket{0}\ket{1}\ket{2}$ obtained by the application of a spin orbital (SO), spin-adapted but spin-incomplete (SASI) and spin-adapted and spin-complete (SASC) $\hat{T}_1$ operator to the doublet reference $\ket{0\overline{0}1}$.}
\end{figure}
\end{center}

It is well known (see e.g. \cite{Stanton1994}) that spin orbital implementations of open-shell CC may introduce spin contaminations into the CC wave function even if they are applied to references being proper spin eigenfunctions (e.g. ROHF references). 
Through the incorporation of spin-adapted (also referred to as spin-free) substitution operators, all contaminations of the latter type can be removed. This leads to pure spin states being exact $\hat{S}^2$ eigenfunctions. In terms of CI or CC calculations e.g., every substitution operator $\hat{T}$ for which 
\begin{equation}
\left[\hat{S}^2, \hat{T}\right] = 0
\end{equation}
holds, conserves the total spin and therefore produces exact spin eigenfunctions (implied that the reference is a spin eigenfunction). Any such $\hat{T}$ may be called spin-adapted. Spin adaption on itself does however not guarantee a sufficiently spanned spin space (i.e. spin completeness). 
Consider, e.g., the spatial single substitution $\ket{0}\ket{0}\ket{1} \rightarrow \ket{0}\ket{1}\ket{2}$, where a particle occupying spatial orbital $0$ is moved to spatial orbital $2$. In the doublet case, a suitable reference determinant is given by $\ket{0\overline{0}1}$ (over-lined indices shall denote $\beta$ electrons while not over-lined indices shall denote $\alpha$ electrons). In this case, the three determinants $\ket{\overline{0}12}$, $\ket{0\overline{1}2}$ and $\ket{01\overline{2}}$ are required to span the complete spin space of the desired spatial configuration $\ket{0}\ket{1}\ket{2}$. As illustrated on the left of Figure \ref{SOSISC.fig}, a standard spin orbital $\hat{T}_1$ operator can however only reach determinants $\ket{\overline{0}12}$ or $\ket{01\overline{2}}$ with distinct amplitudes $t_{0\alpha}^{2\alpha}$ and $t_{0\beta}^{2\beta}$, respectively. Therefore, the space spanned by $\hat{T}_1\ket{0\overline{0}1}$ is spin-incomplete (one determinant is missing) and possibly spin-contaminated for any deviation of the amplitudes $t_{0\alpha}^{2\alpha}$ and $t_{0\beta}^{2\beta}$.

In contrast, the application of a spin-adapted, e.g. purely spatial, $\hat{T}_1$ operator (as illustrated in the center of Figure \ref{SOSISC.fig}) leads to a CSF corresponding to a true $\hat{S}^2$ eigenfunction. A second CSF including the determinant $\ket{0\overline{1}2}$ however, is still missing if creator and annihilator spaces are not allowed to overlap. As illustrated on the right-hand side of Figure \ref{SOSISC.fig}, the second CSF may be recovered by a $0\rightarrow 1\rightarrow 2$ substitution employing the spectator index 1.

As outlined in the following subsections,  the aim for open-shell coupled cluster should always be to reach spin adaption and spin completeness. The usage of spin-adapted but spin-incomplete operators may lead to errors in the final wave function.

In terms of open-shell coupled cluster, a clear distinction can be made for substitution operators incorporating
\begin{itemize}
\item[(I)] spin orbitals and 
\item[(II)] spatial orbitals. 
\end{itemize}

For (I), one of the first contributions to the field of open-shell CC was made by Mukherjee et al. \cite{Mukherjee1975} using spin orbital (i.e. non-spin-adapted) operators at the cost of spin contamination. Later, Szalay et al. \cite{Szalay1997} developed a spin-restricted scheme, which ensures the correct $\hat{S}^2$ expectation value of the wavefunction $e^{\hat{T}}|\Psi_0\rangle$ by the inclusion of spin equations. This leads to exact eigenfunctions of $\hat{S}^2$ only, if spin equations for all CSFs are solved. Usage of the full CSF basis (for every possible $S$) however is only feasible in the simplest cases such that the imposed spin constraints are only followed in a truncated manifold. This however, does not provide a rigorously spin-adapted CC wave function. A comparison between

\onecolumngrid
\twocolumngrid

 spin-restricted and spin-adapted implementations was given by Heckert et al. \cite{Heckert2006}. Furthermore, explicitly correlated and spin-restricted R12 CCSD was developed by Wilke et al. \cite{Wilke2011}  

In type (II) approaches, the cluster operator itself is defined such that it commutes with the $\hat{S}^2$ operator. One of the first mentions of such a symmetry-adapted cluster operator was made by Nakatsuji et al. in the late 70s \cite{Nakatsuji1978, Nakatsuji1978a}. In the early 90s Janssen et al. \cite{Janssen1991} derived CCSD equations for high spin open-shell references explicitly pointing out that spectating indices are missing in the work of Nakatsuji et al. These spectators lead to non-vanishing $\contraction{}{\hat{T}}{}{\hat{H}}\hat{T}\hat{H}$ and $\contraction{}{\hat{T}}{}{\hat{T}}\hat{T}\hat{T}$ contractions in the Baker-Campbell-Hausdorff expansion of $e^{-\hat{T}}\hat{H}e^{\hat{T}}$. The ansatz of Janssen et al. for the cluster operator is however limited to doublet spin states with a single unpaired electron since double spectating operators (e.g. $\hat{E}_{ijvw}^{vwab}$) required for spin completeness in triplet and higher cases are missing. In the following year Neogr\'{a}dy et al. developed a linear CCSD \cite{Neogrady1992} specifically spin-adapted for doublet states. They augmented their implementation to the non-linear CCSD terms \cite{Neogrady1994} as well as to a non-iterative correction for the triples \cite{Neogrady1995}. 

A different ansatz for the spin-adapted cluster operator was followed by Li and Paldus\cite{Li1993, Li1994, Li1995, Li1995a, Li1995b} as well as Jeziorski, Paldus and Jankowski\cite{Jeziorski1995, Jankowski1999}, who used machinery from unitary group theory to derive linear combinations of spatial substitutions, which lead to orthonormal CSFs when applied to the reference CSF. They derived equations for linear CCSD \cite{Li1993} as well as full CCSD \cite{Li1994} with special emphasis on simple doublet, triplet as well as open-shell singlet cases. Their method was applied to the open-shell singlet state of the ozone molecule \cite{Li1995} and extensively analysed for ROHF doublet \cite{Li1995a} as well as open-shell singlet and triplet \cite{Li1995b} instabilities. A further application of orthogonally spin-adapted CCSD was given by Jankowski et al. with respect to van-der-Wals interactions \cite{Jankowski1999}.

Already in 1978 Lindgren \cite{Lindgren1978} suggested to approximate the wave operator $e^{\hat{T}}$ by the normal-ordered $\left\{e^{\hat{T}}\right\}$ to avoid contractions among different cluster operators. Recently this approximation was employed by Nooijen et al. \cite{Nooijen1996, Nooijen2001} for the single-reference high spin open-shell CCSD for doublet spin states as well as by Sen et al. \cite{Sen2012} for the spin-adapted state-universal MRCCSD ansatz.

Usually, normal-order is applied in CC theory as a mathematical tool to derive working equations, which turned out identically without normal-order. Assuming the wave operator itself to be normal-ordered however, neglects any non-vanishing contractions in the latter. This may lead to unknown implications on the wave function.  

Recently, Datta et al. developed a spin-adapted combinatoric open-shell CC (COSCC) for single reference CCSD \cite{Datta2008} and state-universal MRCCSD \cite{Datta2009}. This ansatz assumes a normal-ordered wave operator and reintroduces contractions of spectating substitutions via a combinatoric cluster expansion. Quite recently Datta et al. \cite{Datta2013} developed an automated implementation of COSCCSD for doublet spin states.

Current implementations of rigorously spin-adapted CC methods seem to be limited by the doubles truncation and the triplet spin state. From our perspective this is due to two reasons:
\begin{itemize}
\item[(1)] the appearance of spectating substitutions in the cluster operator leads to non-vanishing contractions among different cluster operators such that the BCH series does not truncate algebraically.
\item[(2)] spatial substitution operator sets beyond the doubles truncation are in the general definition increasingly linearly dependent and therefore bear a certain degree of ambiguity. They also need to be reformulated for every quantum number $S$.
\end{itemize}

While (1) is merely complicating the derivation of working equations itself, (2) is in our opinion a greater challenge to overcome. In this work, we aim at presenting a rigorous scheme to derive linearly independent and spin-complete spin-adapted cluster operators of arbitrary spatial substitution order for arbitrary high spin states. In comparison to the cluster operators of the COSCC \cite{Datta2008, Datta2009, Datta2013} or the orthogonally spin-adapted CC \cite{Li1993, Li1994, Li1995, Li1995a, Li1995b, Jeziorski1995, Jankowski1999} approaches, our cluster operator possesses a much simpler form being composed of solitary spatial substitution operators only. Through the application of L\"owdin's method, our generated cluster operators are linearly independent and spin-complete throughout all truncation levels without the need for specially crafted linear combinations. This may simplify any computer-aided equation derivation and implementation processes needed to arrive at an efficient open-shell CC implementation.

\section{Theory}
\label{Motivation}

In this section, a brief theoretical overview to the operator generation scheme is given. After general definitions of the terms spin adaption and spin completeness in subsection \ref{SpinAdaptionAndSpinCompleteness}, subsection \ref{Problem} defines the linear dependence as well as the spin completeness problem for general (single) spatial substitution operators. In the following two subsections \ref{TheProjectionOperatorMethod} and \ref{LoewdinsMethod}, L\"owdin's projection operator method to generate linearly independent and complete sets of spin eigenfunctions is briefly introduced. Finally, in subsection \ref{ApplicationtoGeneralEOperators} a short motivation on how to use L\"owdin's method to generate spin-complete and linearly independent sets of spatial substitution operators in the closed-shell as well as the more general high spin open-shell case is given.  

\subsection{Spin Adaption and Spin Completeness}
\label{SpinAdaptionAndSpinCompleteness}
Given the Schr\"odinger equation,
\begin{equation}
\Op H |\Psi\> = E |\Psi\>\,,
\end{equation}
in the non-relativistic framework it is 
\begin{equation}
\label{Eq:SS_Sz:Commutation}
[\Op H, \Op S_z] = [\Op H, \Op S^2] = 0.
\end{equation}
Here, $|\Psi\>$ is an element of the linear space $V(n, S, S_z)$
with $n$, $S$, $S_z$ denoting the number of electrons, the total spin quantum number and
the $S_z$ spin projection, respectively. 
Introducing a many-particle basis $\{|\Phi_I\>\}$
the total wavefunction $\ket{\Psi}$ may be represented as
\begin{equation}
|\Psi\> = \sum_I c_I |\Phi_I\>
\end{equation}
with $c_I \in \mathbb{C}$.
Now \eqref{Eq:SS_Sz:Commutation} implies that an individual $|\Phi_I\>$ may 
be written in terms of the symmetric group approach \cite{DuchKarwowski1985} as
\begin{align}
|\Phi_I\> &= \xi_{\lambda_I} \mathcal{A} [|\lambda_I\> \,|O_{\lambda_I}, S, S_z, \nu_I\>]
\end{align}
with $\xi_{\lambda_I} \in \mathbb R$ a scalar prefactor,
$|\lambda_I\>$ the spatial part of $|\Phi_I\>$ ($\to$ configuration),
$O_{\lambda_I}$ the number of open shells of configuration $|\lambda_I\>$, and
$|O, S, S_z, \nu\>$ the spin eigenfunctions fulfilling
\begin{align}
\label{Eq:Sz_Eigenfunctions}
\Op S_z |O, S, S_z, \nu\> &= S_z |O, S, S_z, \nu\>\\
\label{Eq:SS_Eigenfunctions}
\Op S^2 |O, S, S_z, \nu\> &= S(S+1) |O, S, S_z, \nu\>
\end{align}
with $\nu \in \{1\ldots f(O, S)\}$ and $f(O, S)$ the spin eigenfunction degeneracy
depending on the number of open shells $O$ (depending on $|\lambda\>$)
and the total spin quantum number $S$.
The degeneracy of $\Op S^2$ requires special attention as it
increases with the number of open shells in a configuration
due to anti-symmetry.
While ensuring an approximate wavefunction $|\tilde \Psi\>$ to be an eigenfunction of $\Op S_z$ is trivial,
the matter is somewhat more involved for $\Op S^2$  particularly
in the open-shell case and CC approaches.

The introduction of the previous notation now allows for the two following concise definitions:
\begin{enumerate}
\item An approximate wavefunction $|\tilde \Psi\>$ for a given $S$ and $S_z$ is called 
\textit{spin-adapted} if 
\begin{align}
\label{Eq:SpinAdapted}
|\tilde \Psi\> &\in V(n, S, S_z).
\end{align}

\item A linear space $\tilde V$ spanned by a basis 
$\tilde B=\{|\tilde \Phi_I\>\}$ of Slater determinants
is called \textit{spin-complete} w.r.t. $S$ and $S_z$ if 
\begin{multline}
\label{Eq:SpinComplete}
\ls \{|\tilde \Phi_I\> \; | \; |\tilde \lambda_I\> = |\Lambda\>\} = \ls \{\mathcal{A}[|\Lambda\> \,|O_{\Lambda}, S, S_z, \nu\>]\}, \\
\forall_{|\Lambda\> \in \tilde B},\; \nu \in {1\ldots f(O_{\Lambda}, S)}
\end{multline}
with $|\Lambda\> \in \tilde B$ iterating over all configurations in $\tilde B$.
\end{enumerate}
Equation \eqref{Eq:SpinAdapted} states that there must be no components 
from other $S_z$ and $S$ quantum numbers in the wavefunction
while equation
\eqref{Eq:SpinComplete} states that for each configuration appearing
in $\tilde B$ there is a complete set of spin eigenfunctions for that
particular $S_z$ and $S$ which may be freely linearly combined.

In the Configuration Interaction (CI) framework, the approximate wavefunction is usually constructed as
\begin{equation}
\label{Eq:CI}
|\Psi_\text{CI}\> = \sum_I c_I |\Phi_I\>
\end{equation}
which is readily spin-adapted and spin-complete when chosing $\{|\Phi_I\>\}$ 
to span $V(n, S, S_z)$.

An alternative way of constructing the CI wavefunction in intermediate normalization 
in terms of substitution operators reads
\begin{equation}
\label{Eq:CISubst}
|\Psi_\text{CI}\> = (1 + \Op T) |\Phi_0\>
\end{equation}
introducing the substitution
operator $\Op T$ (here we do not explicitly distinguish $\Op C$ and
and $\Op T$ for CI and CC, respectively) containing weighted particle substitutions.

Assuming $|\Phi_0\>$ to be spin-adapted it is a matter 
of $\Op T$ if the resulting $|\Psi_\text{CI}\>$ is spin-adapted and spin-complete
(for truncated $\Op T$).
Spin adaption of $|\Psi_\text{CI}\>$ is readily achieved by chosing
$\Op T$ itself to be spin quantum number conserving, that is
\begin{equation}
\label{Eq:C_SS_Sz:Commutation}
[\Op T, \Op S_z] = [\Op T, \Op S^2] = 0.
\end{equation}
The latter is most easily achieved using 
spatial substitution operators ($\Op E$ operators) making no reference to the spin.
If applied to a closed-shell reference this ansatz
(although linearly dependent starting with triples) is also spin-complete.

In contrast to CI, the CC ansatz refers to substitution operators 
and products thereof explicitly. Its wavefunction $|\Psi_\textrm{CC}\>$ is given by
\begin{equation}
|\Psi_\text{CC}\> = e^{\Op T} |\Phi_0\>
\end{equation}
with $|\Phi_0\>$ the reference (zeroth order wavefunction) and the cluster operator $\Op T$.
The previous discussion for CI w.r.t. to the choice of $\Op T$ is still valid. 
However, for CC the use of substitution operators is mandatory
while for CI it was optional.
Additionally, spin completeness does apply to (truncated) CC
wavefunctions within the non-product space only.
The restriction to non-product terms is necessary here because the 
number of open shells and their associated spin degeneracy
may grow from product terms. That is: the configuration
generated by a product substitution may require more spin eigenfunctions
than the product of the number of eigenfunctions generated by the factor substitutions provides.
In other words: spin completeness for (truncated) CC does 
apply to its linearized part only. Nevertheless, it is still important to ensure.

Spin incompleteness constitutes a significant
restriction of function space for a given configuration (i.e. at a particular spatial substitution level)
and is expected to have a significant impact on the correlation energy. Spin orbital CC is usually truncated at a substitution order of two. This is because, as e.g. stated in\cite{Knowles1993}, the spanned SD manifold $(\hat{T}_1+\hat{T}_2)\ket{\Psi_0}$ reflects all functions interacting with the two-particle-interrelating terms of the Hamiltonian. When conducting a spin-adapted CC using a Hamiltonian and a cluster operator composed of spatial substitutions instead, the same arguments hold for the spatial SD manifold. This means, spatial substitutions such as $\hat{E}_{ijv}^{vab}$, which possess a nominal spatial substitution order of two, hold a direct non-vanishing contribution to the correlation energy and should not be omitted as e.g. in spin-incomplete theories. 
Therefore, the aim for open-shell CC should always be to reach spin adaption and spin completeness.

\subsection{Linear Dependency of Spatial Substitutions}
\label{Problem}

A simple definition of spin-adapted substitution operators $\hat{E}_{r\ldots s}^{p\ldots q}$ is given by the spin integration of spin orbital substitution operators $\hat{X}_{r\sigma_1\ldots s\sigma_m}^{p\sigma_1\ldots q\sigma_m}$ via 
\begin{align}
\hat{E}_{r\ldots s}^{p\ldots q} &= \sum_{\sigma_1 = \alpha, \beta}\cdots\sum_{\sigma_m = \alpha, \beta} \hat{X}_{r\sigma_1\ldots s\sigma_m}^{p\sigma_1\ldots q\sigma_m} \\
&= \sum_{\sigma_1 = \alpha, \beta} \hat{a}_{p\sigma_1}^{\dagger} \left( \ldots \left( \sum_{\sigma_m = \alpha, \beta} \hat{a}_{q\sigma_{m}}^{\dagger}\hat{a}_{s\sigma_{m}} \right) \ldots \right) \hat{a}_{r\sigma_1}\;,
\end{align}

where $p\ldots q$ and $r\ldots s$ denote spatial orbital indices with spin indices $\sigma_1$ to $\sigma_m$. Through index permutations of the creators ($p\ldots q$) and/or the annihilators ($r\ldots s$), all possible $\hat{E}$ operator compositions may be obtained. In this work, we define $\hat{E}$ operator index permutations via a permutation vector of two components. When applied to an $\hat{E}$ operator, the upper component acts on the creators while the lower component acts on the annihilators via
\begin{equation}
\begin{pmatrix} \hat{P} \\ \hat{P}' \end{pmatrix} \hat{E}_{r\ldots s}^{p\ldots q} = \hat{E}_{\hat{P}' (r\ldots s)}^{\hat{P} (p\ldots q)}\quad \forall_{\hat{P},\hat{P}' \in \mathbb{S}_m}
\end{equation}
with $\mathbb{S}_m$ being the symmetric group of order $m$. In total there are $(m!)^2$ possible $\hat{E}$ operator compositions.  
Due to the pair-wise spin summation, only the relative ordering of annihilator ($r\ldots s$) and creator ($p\ldots q$) indices matters such that $m!$ operator compositions are trivially identical with
\begin{equation}
\hat{E}_{r\ldots s}^{p\ldots q} = \begin{pmatrix} \hat{P} \\ \hat{P}' \end{pmatrix} \hat{E}_{r\ldots s}^{p\ldots q} \quad \forall_{\hat{P} = \hat{P}'}.
\end{equation} 

Therefore, the remaining $m!$ non-trivial $\hat{E}$ operators are e.g. given by 
\begin{equation}
\bigcup_{\hat{P} \in \mathbb{S}_m} \hat{E}_{r\le\ldots\le s}^{\hat{P}(p\ldots q)} \text{ or } \bigcup_{\hat{P} \in \mathbb{S}_m} \hat{E}_{\hat{P} (r\ldots s)}^{p\le\ldots \le q},
\end{equation}
where either the annihilators or the creators are sorted in ascending order.

For any spatial substitution $\hat{E}$, the number of linearly independent operators, i.e. the minimal amount of operators to span the complete spin space, is given by the spin degeneracy $f(O, S)$ (see e.g. \cite{Pauncz1979}). The latter depends on 
the number of open shells $O$ as well as the spin quantum number $S$ with 
\begin{equation}
\label{SpinDegeneracy.eq}
f(O, S) = \binom{O}{\frac{O}{2} - S} - \binom{O}{\frac{O}{2} - S - 1}\le m!\,.
\end{equation}

In general, the number of possible permutations $m!$ is larger than the number of linearly independent operators $f(O, S)$. If the linearly dependent operators are not removed, the spanned CI or CC wave function is overparametrized. This also produces linear dependencies in the residual equations. Please note that closed-shell CCSD or CISD resemble special cases, where $f(O,S)$ is identical to $m!$ simplifying their treatment and implementation substantially.  

In this work we intent to introduce a routine to systematically derive all permutations $\hat{P}$ leading to linearly independent but full spin-space-spanning (i.e. spin-complete) $\hat{E}$ operators. 

\subsection{The Projection Operator Method}
\label{TheProjectionOperatorMethod}

Several techniques to generate spin eigenfunctions are known in the literature (see e.g. \cite{Pauncz1979}). In contrast to 
iterative procedures, where the spin eigenfunctions of smaller
subsystems are expanded using the addition theorem of
angular momenta, the projection operator method, originally introduced by L\"owdin \cite{Lowdin1955, Lowdin1964}, eliminates all components of an arbitrary trial function (in the space spanned by the spin eigenfunctions), which do not correspond to the correct $S(S+1)$ eigenvalue.

When applied to arbitrary trial functions $\theta_i$, the operator $\hat{S}^2 - K(K+1)$ will annihilate all spin eigenfunctions of spin quantum number $K$. To only keep components of a specific spin quantum number $S$, the product of annihilation operators is used: 
\begin{equation}
\hat{O}_S = \prod_{K \neq S}\frac{\hat{S}^2 - K(K+1)}{S(S+1) - K(K+1)}
\end{equation}
The denominator ensures that correct spin eigenfunctions with spin quantum number $S$ are unchanged. 
It is convenient to use $\hat{S}_z$ eigenfunctions as trial functions.
Consider the sorted primitive $\hat{S}_z$ eigenfunction $\theta_1$ with 
\begin{align}
\theta_1 &= \alpha(1)\ldots\alpha(\mu)\beta(\mu + 1)\ldots\beta(\mu + \nu)\,, \\
\hat{S}_z\theta_1 &= \frac{1}{2}(\mu - \nu)\theta_1 = S_z\theta_1\;.
\end{align}

It was shown \cite{Lowdin1955} that the projected spin eigenfunction $\Theta_1$ for the high spin case $S=S_z$ gained from the application of $\hat{O}_{S=S_z}$ to $\theta_1$ is given by 
\begin{align}
\Theta_1 &= \hat{O}_{S=S_z}\theta_1 \\
&= \frac{2S+1}{\mu + 1}\sum_{q=0}^{\nu}(-1)^q\binom{\mu}{q}^{-1}\left[\alpha^{\mu-q}\beta^{q}\right]\left[\alpha^{q}\beta^{\nu-q}\right]\;,
\label{Theta_1.eq}
\end{align}
where square brackets $\left[\alpha^{a}\beta^{b}\right]$ are used to denote the sum of all possible primitive spin functions with $a$ $\alpha$-functions and $b$ $\beta$-functions. As an example consider the term $\left[\alpha^2\beta\right]$ with 
\begin{equation*}
\left[\alpha^2\beta\right] = \alpha(1)\alpha(2)\beta(3) + \alpha(1)\beta(2)\alpha(3) + \beta(1)\alpha(2)\alpha(3).
\end{equation*}
For a given $S_z$ quantum number, all primitive $\hat{S}_z$ eigenfunctions possess the same number of $\alpha$ particles and the same number of $\beta$ particles. Therefore, they are connected by simple particle permutations. 
For a given set of $n$ primitive $\hat{S}_z$ eigenfunctions $\left\{\theta_1,\ldots ,\theta_n\right\}$, we define the permutation operator $\hat{P}_i^j$ to relate eigenfunctions $\theta_i$ and $\theta_j$ via 
\begin{equation*}
\hat{P}_i^j \theta_i = \theta_j.
\end{equation*}

Due to the fact that arbitrary particle permutations $\hat{P}_{i}^{j}$ commute with the projection operator $\hat{O}_{S=S_z}$ \cite{Pauncz1979}, the projection of an arbitrary primitive spin function $\theta_i$ can be written as 
\begin{equation}
\label{Theta_i.eq}
\Theta_i = \hat{O}_{S=S_z}\theta_i = \hat{O}_{S=S_z}\hat{P}_{1}^{i}\theta_1 = \hat{P}_{1}^{i}\Theta_1\;.
\end{equation}

It was shown \cite{Gershgorn1968} that the complete set of (non-orthogonal, high spin) spin eigenfunctions $\Theta_i$ is obtainable via these particle permutations. Once a single spin eigenfunction is found (e.g. $\Theta_1$ via equation \ref{Theta_1.eq}), all other spin eigenfunctions may be gained through particle permutations.

\subsection{L\"owdin's Method to Select Linearly Independent Eigenfunctions}
\label{LoewdinsMethod}

With increasing number of particles, the total number of particle permutations $\hat{P}_{i}^{j}$ may get larger than the number of linearly independent 
spin eigenfunctions. As one is generally interested in the linearly independent spin eigenfunctions only, L\"owdin developed a scheme \cite{Lowdin1955, Lowdin1964} 
to select only those permutations, which 
lead to linearly independent and complete spin eigenfunctions $\Theta_i$, which was later proven by Gershgorn \cite{Gershgorn1968}.

\begin{center}
\begin{figure}[h]
\includegraphics[width=\textwidth]{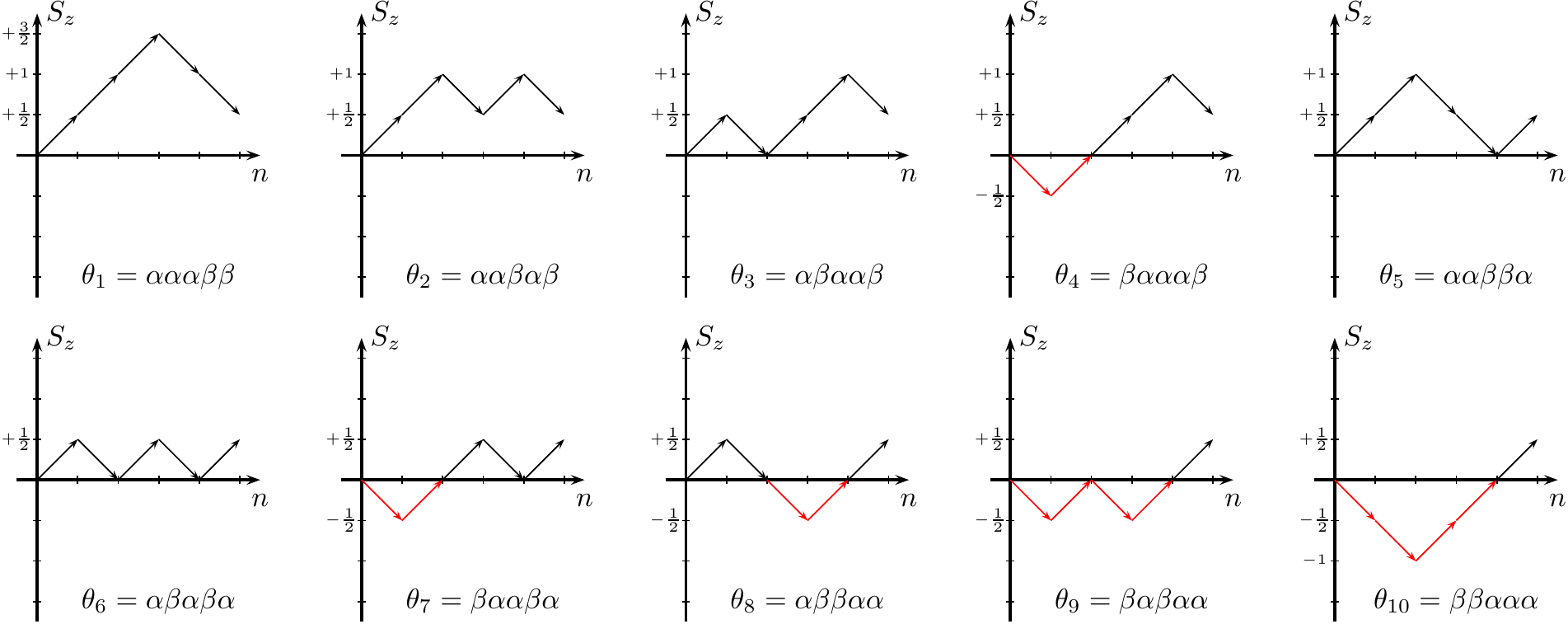}
\caption{\label{Loewdin.fig}Path diagrams for all primitive spin functions for $n=5$ and $S_z = \frac{1}{2}$. Line segments below the $S_z = 0$ reference line are displayed in red.}
\end{figure}
\end{center}

Any primitive $\hat{S}_z$ eigenfunction $\theta_i$ can be visualized as a distinct path diagram, where each $\alpha$ corresponds to a line segment pointing in the direction of $+45^{\circ}$ (up) and each $\beta$ to a line segment pointing in the direction of $-45^{\circ}$ (down).
Consider e.g. all primitive spin functions for $n = 5$ and $S_z = +\frac{1}{2}$ as displayed in Figure \ref{Loewdin.fig}.

\onecolumngrid 
\twocolumngrid

For the high spin case ($S = S_z$), the collection of path diagrams whose line segments lie entirely above the $S_z = 0$ reference axis lead to linearly independent spin eigenfunctions. In the given example (Figure \ref{Loewdin.fig}), primitive spin functions $\theta_1$ -- $\theta_3$ and $\theta_5$ -- $\theta_6$ show this behaviour. According to equation (\ref{Theta_i.eq}), the permutations shown in table \ref{LinearIndependentPermutations.tab} lead to a set of linearly independent spin eigenfunctions when applied to $\Theta_1$. Individual particle transpositions of particles $x$ and $y$ are denoted by $(x , y)$.
 
\begin{table}[h]
\caption{\label{LinearIndependentPermutations.tab} Permutations $\hat{P}_1^i$ leading to linearly independent spin eigenfunctions for $n = 5$ and $S = S_z = \frac{1}{2}$.}
\begin{tabular}{c | c}
\textbf{Primitive spin function} & \textbf{Permutation $\hat{P}_1^i$} \\ \hline
$\theta_1 = \alpha\alpha\alpha\beta\beta$ & $\hat{P}_1^1 = \hat{1}\phantom{(, 4)}$ \\
$\theta_2 = \alpha\alpha\beta\alpha\beta$ & $\hat{P}_1^2 = (3, 4)$ \\
$\theta_3 = \alpha\beta\alpha\alpha\beta$ & $\hat{P}_1^3 = (2, 4)$ \\
$\theta_5 = \alpha\alpha\beta\beta\alpha$ & $\hat{P}_1^5 = (3, 5)$ \\
$\theta_6 = \alpha\beta\alpha\beta\alpha$ & $\hat{P}_1^6 = (2, 5)$
\end{tabular}
\end{table}

This leads to the complete basis $\mathcal{B}$ of spin eigenfunctions with
\begin{equation*}
\mathcal{B} = \left\{\begin{matrix}\phantom{\hat{P}_1^2}\Theta_1 \\ \hat{P}_1^2\Theta_1 \\ \hat{P}_1^3\Theta_1 \\ \hat{P}_1^5\Theta_1 \\ \hat{P}_1^6\Theta_1 \end{matrix}\right\}\,.
\end{equation*}

\subsection{Application to General $\hat{E}$ Operators}
\label{ApplicationtoGeneralEOperators}

In the following two subsections the application of L\"owdin's method to general types of $\hat{E}$ operators to systematically derive linear independent 
index permutations is developed. After discussing the simpler closed-shell case, the method is generalized to arbitrary high spin open-shell systems. 

\subsubsection{Closed-Shell Reference}
\label{ClosedShellReference}

For arbitrary closed-shell systems, the common index notation for occupied $\mathbb{O}$, virtual $\mathbb{V}$ and joined $\mathbb{O} \cup \mathbb{V}$ spatial orbital spaces is used: 
\begin{align*}
i, j, k, \ldots &\in \mathbb{O} \\
a, b, c, \ldots &\in \mathbb{V} \\
p, q, r, \ldots &\in \mathbb{O} \cup \mathbb{V}
\end{align*}

Any closed-shell reference determinant for $\mu$ $\alpha$ and $\nu$ $\beta$ electrons with $\mu = \nu$ may be written as 
\begin{equation}
\label{ClosedShellReference.eq}
\ket{\Psi_0} = \ket{i_1\overline{i_1}\ldots i_{\mu}\overline{i_{\mu}}} = \sqrt{n!}\hat{\mathcal{A}} \left[ \Phi(1\ldots n) \Theta(1\ldots n) \right]\,,
\end{equation}

where over-lined indices denote $\beta$ electrons and not over-lined indices $\alpha$ electrons, $n=2\mu$ denotes the number of particles, $\Phi$ the spatial part of the determinant, $\Theta$ its spin part (the spin eigenfunction) and $\hat{\mathcal{A}}$ denotes the antisymmetrizer with 
\begin{equation}
\hat{\mathcal{A}} = \frac{1}{n!}\sum_{\hat{P}\in\mathbb{S}_n} (-1)^{p(\hat{P})} \hat{P}\,,
\end{equation}

where $p(\hat{P})$ denotes the parity of permutation $\hat{P}$. In general, particle permutations $\hat{P}$ effect both spatial and spin parts such that 
\begin{equation}
\label{permutationspinspatial.eq}
\hat{P} = \hat{P}^{\Phi}\hat{P}^{\Theta}
\end{equation}
where $\hat{P}^{\Phi}$ acts on the spatial part $\Phi$ and $\hat{P}^{\Theta}$ on the spin part $\Theta$, only.

Due to the abundance of open shells in $\ket{\Psi_0}$ and the spin quantum number $S = 0$, the corresponding configurational spin space is singly degenerate since 
\begin{equation*}
f(0,0) = \binom{0}{0} - \binom{0}{-1} = 1\,.
\end{equation*}

Therefore, all spin eigenfunctions $\Theta_i$ from projected primitive spin functions $\theta_i$ (in the correct $Sz = 0$ space) lead to linearly dependent reference determinants when combined with the closed-shell spatial part $\Phi$ and antisymmetrized. For convenience, we choose $\Theta$ to be 
\begin{equation}
\label{ConvenientChoice.eq}
\Theta = \hat{O}_{S=S_z=0} \alpha(1)\beta(2)\ldots\alpha(n-1)\beta(n)\,.
\end{equation}

Arbitrary spatial substitutions $\hat{E}_{i_1 \le \ldots\le i_m}^{a_1\ldots a_m}$ applied to (\ref{ClosedShellReference.eq}) do not act on the spin part $\Theta$. Since they commute with the antisymmetrizer \cite{Pauncz1979}, they will only affect the spatial part $\Phi$ with
\begin{align}
\notag 
\hat{E}_{i_1 \le \ldots\le i_m}^{a_1\ldots a_m}\ket{\Psi_0} &= \sqrt{n!}\hat{\mathcal{A}}\left[\left(\hat{E}_{i_1 \le \ldots\le i_m}^{a_1\ldots a_m}\Phi(1\ldots n)\right)\right. \\ 
&\left.\phantom{==} \Theta(1\ldots n)\right]\,,
\end{align}
At this point, several closed shells of $\ket{\Psi_0}$ may be opened (through the application of $\hat{E}$ to $\Phi$). Despite any changes to the actual spin eigenfunction $\Theta$, the latter will adapt to the new spatial part by possibly creating different spin eigenfunctions in the total CSF (through means of the antisymmetrizer). 

From L\"owdin's method (subsection \ref{LoewdinsMethod}) all particle permutations leading to linearly independent spin eigenfunctions for a given number of created open shells are known. In the following, these permutations are denoted by $\hat{P}_{(i)}^{\Theta}$. For the given spatial substitution $\hat{E}_{i_1 \le \ldots\le i_m}^{a_1\ldots a_m}$, which leads to a total of $O$ open shells, the $f = f(O,0)$ necessary linearly independent CSFs are therefore given by 
\begin{align*}
\ket{\Psi_{i_1\ldots i_m}^{a_1\ldots a_m}}^{(1)} &= \sqrt{n!}\hat{\mathcal{A}}\left[\left(\hat{E}_{i_1\le\ldots\le i_m}^{a_1\ldots a_m}\Phi\right)\Theta\right] \\
\ket{\Psi_{i_1\ldots i_m}^{a_1\ldots a_m}}^{(2)} &= \sqrt{n!}\hat{\mathcal{A}}\left[\left(\hat{E}_{i_1\le\ldots\le i_m}^{a_1\ldots a_m}\Phi\right)\hat{P}_{(2)}^{\Theta}\Theta\right] \\
&\,\,\,\vdots \\
\ket{\Psi_{i_1\ldots i_m}^{a_1\ldots a_m}}^{(f)} &= \sqrt{n!}\hat{\mathcal{A}}\left[\left(\hat{E}_{i_1\le\ldots\le i_m}^{a_1\ldots a_m}\Phi\right)\hat{P}_{(f)}^{\Theta}\Theta\right]\,.
\end{align*}

Using equation (\ref{permutationspinspatial.eq}) for each CSF $\ket{\Psi_{i_1\ldots i_m}^{a_1\ldots a_m}}^{(i)}$, it is

\begin{align*}
\ket{\Psi_{i_1\ldots i_m}^{a_1\ldots a_m}}^{(i)} &= \sqrt{n!}\hat{\mathcal{A}}\left[\left(\hat{E}_{i_1\le\ldots\le i_m}^{a_1\ldots a_m}\Phi\right)\hat{P}_{(i)}^{\Theta}\Theta\right]\\
&= \sqrt{n!}\underbrace{\hat{\mathcal{A}}\hat{P}_{(i)}}_{\hat{\mathcal{A}}}\left[\left(\left(\hat{P}_{(i)}^{\Phi}\right)^{-1}\hat{E}_{i_1\le\ldots\le i_m}^{a_1\ldots a_m}\Phi\right)\Theta\right] \\
&= \sqrt{n!}\hat{\mathcal{A}}\left[\left(\left(\hat{P}_{(i)}^{\Phi}\right)^{-1}\hat{E}_{i_1\le\ldots\le i_m}^{a_1\ldots a_m}\Phi\right)\Theta\right]\,.
\end{align*}

The spin particle permutation $\hat{P}_{(i)}^{\Theta}$ can be expressed by the inverse spatial particle permutation $\left(\hat{P}^{\Phi}_{(i)}\right)^{-1}$ applied to the spatial part $\hat{E}_{i_1\le\ldots\le i_m}^{a_1\ldots a_m}\Phi$. This spatial permutation may be adopted to the $\hat{E}$ operator at hand by either permuting the annihilator or the creator indices. Further details of this technique will be 
given in section \ref{Implementation}. 

As an example consider the two-fold substitution 
\begin{equation*}
\hat{E}_{ij}^{ab} \ket{i\overline{i}j\overline{j}} = \ket{\overline{i}ja\overline{b}} + \ket{i\overline{j}\overline{a}b} -\ket{ij\overline{a}\overline{b}} - \ket{\overline{i}\overline{j}ab}\,,
\end{equation*}
creating four open shells in the final CSF. The corresponding spin degeneracy is $f(4, 0) = 2$. The primitive spin functions whose path diagrams lie entirely above the $S_z = 0$ reference axis are given by 
\begin{equation*}
\left\{\begin{matrix}\theta_1 = \alpha\beta\alpha\beta\phantom{ = (2, 3)\theta_1} \\ \theta_2 = \alpha\alpha\beta\beta = (2, 3)\theta_1\end{matrix}\right\}\,,
\end{equation*}
where $\theta_1$ was arbitrarily chosen to resemble the determinant's short hand notation (\ref{ConvenientChoice.eq}). This leads to the CSFs 
\begin{align*}
\ket{\Psi_{ij}^{ab}}^{(1)} &= \sqrt{4!}\hat{\mathcal{A}}\left[\left(\hat{E}_{ij}^{ab}\ket{i(1)}\ket{i(2)}\ket{j(3)}\ket{j(4)}\right) \Theta_1\right] \\
&= \hat{E}_{ij}^{ab}\ket{i\overline{i}j\overline{j}}\,, \\
\ket{\Psi_{ij}^{ab}}^{(2)} &= \sqrt{4!}\hat{\mathcal{A}}\left[\left(\hat{E}_{ij}^{ab}\ket{i(1)}\ket{i(2)}\ket{j(3)}\ket{j(4)}\right) (2,3)\Theta_1\right] \\
&= \sqrt{4!}\hat{\mathcal{A}}\left[\left((3, 2)\hat{E}_{ij}^{ab}\ket{i(1)}\ket{i(2)}\ket{j(3)}\ket{j(4)}\right) \Theta_1\right]\,.
\end{align*}
The spatial particle permutation $(3, 2)$ is translatable to the spatial orbital index permutation $(j, i)$ through the orbital$\leftrightarrow$particle mapping of the spatial reference configuration $\ket{i(1)}\ket{i(2)}\ket{j(3)}\ket{j(4)}$ and can be absorbed into the $\hat{E}_{ij}^{ab}$ operator to yield 
\begin{align*}
\ket{\Psi_{ij}^{ab}}^{(2)} &= \begin{pmatrix} \hat{1} \\ (j, i) \end{pmatrix}\hat{E}_{ij}^{ab}\ket{i\overline{i}j\overline{j}} = \hat{E}_{ji}^{ab}\ket{i\overline{i}j\overline{j}} \\
&= \ket{ij\overline{a}\overline{b}} + \ket{\overline{i}\overline{j}ab} - \ket{\overline{i}j\overline{a}b} - \ket{i\overline{j}a\overline{b}}\,, 
\end{align*}
which is clearly linearly independent to $\ket{\Psi_{ij}^{ab}}^{(1)}$. Please note that the transformation of spin particle to spatial orbital permutations is not always as trivial as this example might indicate. Detailed rules to this transformation are given in section \ref{Implementation}.

\subsubsection{High Spin Open-Shell Reference}
\label{HighSpinOpenShellReference}

\begin{figure}[h]
\includegraphics[scale=1]{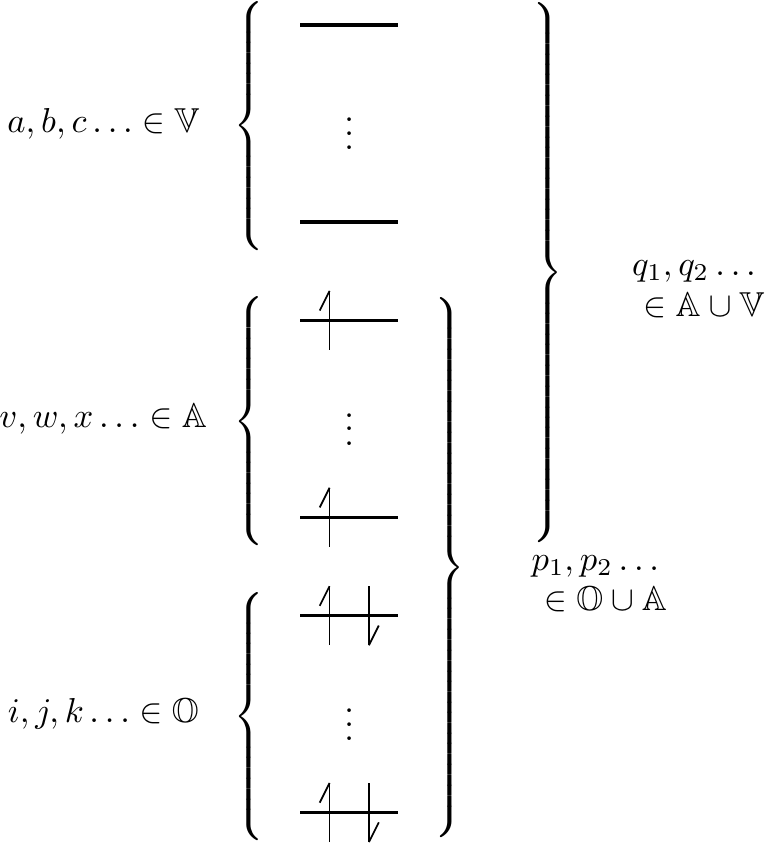}
\caption{\label{HighSpinOrbitalSpace.fig} Index notations for the general $S = S_{z}$ high spin case used in this work.}
\end{figure}

In the general $S = S_z$ high spin open-shell case, index notations according to Figure \ref{HighSpinOrbitalSpace.fig} are used.

To achieve spin completeness, it is mandatory to use overlapping annihilator $p_1,p_2\ldots \in \mathbb{O}\cup\mathbb{A}$ and creator $q_1,q_2\ldots\in\mathbb{A}\cup\mathbb{V}$ spaces in the sense that 
all singly occupied spatial orbitals $v,w,x,\ldots$ can be thought of as elements of an active space $\mathbb{A}$. Arbitrary reference determinants $\ket{\Psi_0}$ of $n_o$ doubly occupied orbitals and $n_a$ singly occupied orbitals (with $\alpha$ electrons) such that $n = 2n_o + n_a$ then take the form 
\begin{align}
\ket{\Psi_0} &= \ket{i_1\overline{i_1}\ldots i_{n_o}\overline{i_{n_o}}v_1\ldots v_{n_a}} \\
&= \sqrt{n!}\hat{\mathcal{A}}\left[\Phi(1\ldots n)\Theta(1\ldots n)\right]\,.
\end{align}
The configurational spin degeneracy is given by 
\begin{equation}
f\left(n_a, \frac{n_a}{2}\right) = \binom{n_a}{0} - \binom{n_a}{-1} = 1\,.
\end{equation}
Following the same argument as in the closed-shell case, we may choose $\Theta$ to be a projected primitive spin function $\theta$ with 
\begin{equation}
\theta = \alpha(1)\beta(2)\ldots\alpha(2n_o-1)\beta(2n_o)\alpha(2n_o+1)\ldots\alpha(n)\,.
\end{equation}
General $\hat{E}_{p_1\le\ldots\le p_m}^{q_1\ldots q_m}$ operators may create additional $\Delta O$ open shells in the $n_a$ open-shell reference. 
The $f = f(n_a+\Delta O, n_a/2)$ linearly independent CSFs are given by 
\begin{align}
\ket{\Psi_{p_1\ldots p_m}^{q_1\ldots q_m}}^{(1)} &= \sqrt{n!}\hat{\mathcal{A}}\left[\left(\hat{E}_{p_1\le\ldots\le p_m}^{q_1\ldots q_m}\Phi\right)\Theta\right] \\
\ket{\Psi_{p_1\ldots p_m}^{q_1\ldots q_m}}^{(2)} &= \sqrt{n!}\hat{\mathcal{A}}\left[\left(\left(\hat{P}^{\Phi}_{(2)}\right)^{-1}\hat{E}_{p_1\le\ldots\le p_m}^{q_1\ldots q_m}\Phi\right)\Theta\right] \\
&\vdots \notag\\
\ket{\Psi_{p_1\ldots p_m}^{q_1\ldots q_m}}^{(f)} &= \sqrt{n!}\hat{\mathcal{A}}\left[\left(\left(\hat{P}^{\Phi}_{(f)}\right)^{-1}\hat{E}_{p_1\le\ldots\le p_m}^{q_1\ldots q_m}\Phi\right)\Theta\right]\,.
\end{align}

Depending on the number of open shells $n_a$ in the reference determinant and the number of additionally created open shells $\Delta O$ of the specific $\hat{E}$ operator, further spectating index pairs might enter the operators via $\left(\hat{P}_{(i)}^{\Phi}\right)^{-1}$. These spectators allow for additional permutational freedom necessary to complete the spin space. In general, for every $\Delta O$ multiple of two (will be generally shown in section \ref{Implementation}), one additional spectator $v_i\rightarrow v_i$ must enter the operator as we shall see in the following example.

Consider the same example substitution $\hat{E}_{ij}^{ab}$ as in the closed-shell case (subsection \ref{ClosedShellReference}) applied to a reference with one open shell ($S = \frac{1}{2}$) via 
\begin{equation}
\hat{E}_{ij}^{ab} \ket{i\overline{i}j\overline{j}v} = + \ket{\overline{i}jva\overline{b}} + \ket{i\overline{j}v\overline{a}b} - \ket{ijv\overline{a}\overline{b}} - \ket{\overline{i}\overline{j}vab}\,.
\end{equation}
In contrast to $f(4,0) = 2$ in the closed-shell case, the spin degeneracy is now given by 
\begin{equation}
f\left(5, \frac{1}{2}\right) = \binom{5}{2} - \binom{5}{1} = 5\,.
\end{equation}
The corresponding primitive spin functions whose path diagrams lie entirely above the $S_z = 0$ reference axis are given by (c.f. Figure \ref{Loewdin.fig})
\begin{equation}
\label{OpenShellExample.eq}
\left\{\begin{matrix}\theta_1 = \alpha\beta\alpha\beta\alpha\phantom{ = (2, 3)(4, 5)\theta_1} \\ 
\theta_2 = \alpha\alpha\beta\beta\alpha = (2, 3)\theta_1\phantom{(2, 3)}\\
\theta_3 = \alpha\beta\alpha\alpha\beta = (4, 5)\theta_1\phantom{(2, 3)}\\
\theta_4 = \alpha\alpha\beta\alpha\beta = (2, 3)(4, 5)\theta_1\\
\theta_5 = \alpha\alpha\alpha\beta\beta = (2, 5)\theta_1\phantom{(2, 3)}\end{matrix}\right\}\,.
\end{equation}

To incorporate the inverse spatial particle permutations of (\ref{OpenShellExample.eq}) into the $\hat{E}_{ij}^{ab}$ operator, the latter permutations need to be transformed to spatial orbital permutations. As in the closed-shell example, this can be done by means of the orbital$\leftrightarrow$particle mapping of the spatial reference configuration $\ket{i(1)}\ket{i(2)}\ket{j(3)}\ket{j(4)}\ket{v(5)}$. Any particle permutation involving particle index five indicates a permutation of the spatial orbital $v$. Since $v$ is not part of the original $\hat{E}_{ij}^{ab}$ operator, it has to be augmented to $\hat{E}_{ijv}^{abv}$. This leads to the final linearly independent $\hat{E}$ operators 
\begin{equation}
\left\{
\begin{matrix}
\phantom{(v, j)}(j, i)\hat{E}_{ijv}^{abv} = \hat{E}_{ij}^{ab} \\
\phantom{(j, i)}(j, i)\hat{E}_{ijv}^{abv} = \hat{E}_{ji}^{ab} \\
\phantom{(j, i)}(v, j)\hat{E}_{ijv}^{abv} = \hat{E}_{ivj}^{abv} \\
(v, j)(j, i)\hat{E}_{ijv}^{abv} = \hat{E}_{vij}^{abv} \\ 
\phantom{(j, i)}(v, i)\hat{E}_{ijv}^{abv} = \hat{E}_{vji}^{abv}
\end{matrix}
\right\}\,.
\end{equation}

\section{Methodology}
\label{Implementation}

In this section, details of the generation of linearly independent spatial substitution operators $\hat{E}$ are given. Following the general structure shown in Figure \ref{GeneralStructure.fig}, all necessary steps to arrive at a final set of operators are outlined in subsections \ref{PrototypeGeneration} to \ref{IndexSymmetry}: 

\begin{figure}[h]
\includegraphics[width = .5\textwidth]{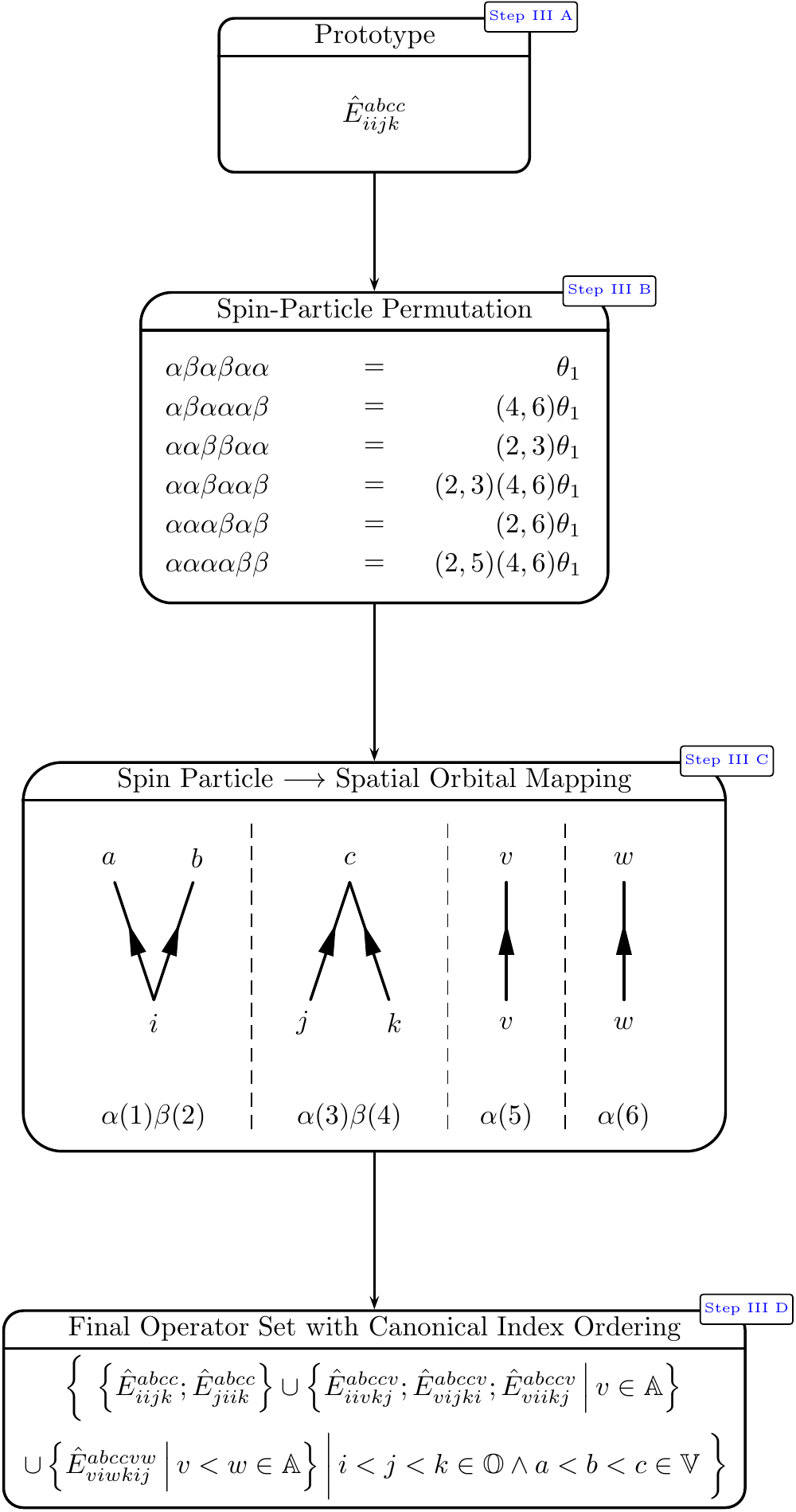}
\caption{\label{GeneralStructure.fig} Flowchart diagram showing the route from operator prototype to final set of linearly independent operators for one example.}
\end{figure}

\begin{itemize}
\item[(\ref{PrototypeGeneration})] All different types of $\hat{E}$ operators for a given substitution rank need to be derived. This includes all operators leading to distinct spatial functions in the resultant CSF to reach spatial completeness. Each different operator type will be denoted by a specific 
$\hat{E}$ prototype. Operators leading to the same spatial function (but e.g. a different spin function) must not be included in the prototype derivation since they will be explicitly derived later.
\item[(\ref{IndexPermutation})]The next step is the generation of spin particle permutations of primitive spin functions that lead to linearly independent spin eigenfunctions when projected according to L\"owdin's method (c.f. \ref{LoewdinsMethod}). 
\item[(\ref{SpinSpatialMapping})]Following the generation of all prototypes and spin particle permutations, explicit mappings from spin particle to spatial orbital permutations need to be given. In this work this mapping problem is redefined as a topological problem.
\item[(\ref{IndexSymmetry})]Finally, the resultant operators need to be gathered into sets with canonical index ordering.
\end{itemize}

\subsection{Prototype Generation}
\label{PrototypeGeneration}

The aim of the $\hat{E}$ prototype generation is to create all possible spatial substitution patterns leading to different spatial functions $\Phi$ for arbitrary high spin references. In this work, an iterative approach was developed, where a known set of $\hat{E}$ prototypes of a certain rank $m$ is augmented 
to a prototype set of rank $m+1$ using explicit index augmentations. Employing the index notation of subsection \ref{HighSpinOpenShellReference}, the initial set $\mathbb{P}_m$ for $m = 1$ consists of the three prototypes 
\begin{equation}
\mathbb{P}_1 = \left\{\hat{E}_{i}^{a}\;;\;\hat{E}_{i}^{v}\;;\;\hat{E}_{v}^{a}\right\}\,,
\end{equation}
containing all single substitutions from $\mathbb{O}$ to $\mathbb{V}$, from $\mathbb{O}$ to $\mathbb{A}$ and from $\mathbb{A}$ to $\mathbb{V}$. These three also represent the possible building blocks for a rank augmentation. If the annihilators of arbitrary $\hat{E}$ operators are assumed to be ordered, the options for allowed index augmentations from the right are limited. An occupied index $i$ in the annihilator space must then e.g. always be followed by (i) the same index $i$, (ii) a higher index $j>i$ or (iii) a new active index $v$, while an active index $v$ in the annihilator space can only be followed by a higher active index $w > v$.\footnote{Here we assumed that in the global index set $\mathbb{O}\cup\mathbb{A}\cup\mathbb{V}$ it is $i < v \quad \forall_{i\in\mathbb{O}, v\in\mathbb{A}}$.} Unfortunately, there is no such restriction on the creator indices. The creator indices represent arbitrary subsets of $\mathbb{A}\cup\mathbb{V}$. 
 
To generate all prototypes leading to distinct spatial functions, the following routine was implemented: 
\begin{itemize}
\item[(a)] Start with an initial set of prototypes $\mathbb{P}_m$ of rank $m$.
\item[(b)] For all operators $\hat{E}$ in $\mathbb{P}_m$:
\begin{itemize}
\item[(i)] Analyze the rightmost annihilator/creator pair of $\hat{E}$ and apply all possible index augmentations to the right while keeping the set of all annihilating indices ordered.
\item[(ii)] Iterate over all permutations of the symmetric group $\mathcal{S}_{m}$ and permute the set of creators accordingly.
\item[(iii)] Check if the augmented and permuted operator is zero. 
\begin{itemize}
\item[\texttt{\textbf{False: }}] proceed to (iv)
\item[\texttt{\textbf{True: }}] continue
\end{itemize}
\item[(iv)] Insert augmented operator into $\mathbb{P}_{m+1}$ if not already contained.
\end{itemize}
\end{itemize}

\subsection{Spin Particle Permutation}
\label{IndexPermutation}

After having generated all spatial substitutions, which lead to a spatial-complete set of  distinct spatial functions, i.e. the $\hat{E}$-prototypes, the actual primitive spin functions that lead to a set of linearly independent spin eigenfunctions when projected according to L\"owdin's method (c.f. subsection \ref{LoewdinsMethod}) need to be generated. 

To achieve spin completeness, a total of $f(O,S)$ (c.f. equation \ref{SpinDegeneracy.eq}) linearly independent CSFs (and therefore $\hat{E}$ operators) per spatial configuration are mandatory. Clearly, this number depends on the number of open shells $O$ (as well as the constant spin quantum number $S$) only. For a given operator prototype, the amount of required spin-complete operators depends on any changes in the total number of open shells $\Delta O$ caused by the application of the latter prototype to the reference determinant. 

In general, the required $f(\Delta O + 2S, S)$ primitive spin functions (above the $S_z = 0$ reference axis) for L\"owdin's method are given for primitive spin functions composed of e.g.
\begin{itemize}
\item[(i)] $\frac{\Delta O}{2}$ $\alpha\beta$ pairs and
\item[(ii)] $2S$ single $\alpha$ particles to set the correct $S = S_z$.
\end{itemize}
Please note that (i) and (ii) can result in primitive spin functions composed of fewer particles than actually contained in the respective reference determinant. This is because only the particles representing additionally opnened shells as well as the particles constituting the spin quantum number $S = S_z$ need to be represented here.

To motivate this, consider for example the prototype $\hat{E}_{ij}^{av}$. This prototype moves two particles from two doubly occupied orbitals $i$ and $j$ to the virtual orbital $a$ as well as the singly occupied orbital $v$. An appropriate high spin reference determinant $\ket{\Psi_0}$ must therefore be composed of at least five particles with 
\begin{equation*}
\ket{\Psi_0} = \ket{\ldots i\overline{i}\ldots j\overline{j} \ldots v\ldots}\,.
\end{equation*}

Considering the simplest (five particle) reference, the application of $\hat{E}_{ij}^{av}$ to the latter leads to 
\begin{equation*}
\hat{E}_{ij}^{av}\ket{i\overline{i}j\overline{j}v} = \ket{ijv\overline{v}\overline{a}} - \ket{\overline{i}jv\overline{v}a}\,,
\end{equation*}
for which $\Delta O = 3-1 = 2$ such that the spin degeneracy is given by 
\begin{equation*}
f\left(2+1, \frac{1}{2}\right) = \binom{3}{1} - \binom{3}{0} = 2\,.
\end{equation*}

Clearly, the five particle primitive spin function $\alpha\beta\alpha\beta\alpha$ resembling the short hand notation of the reference determinant, would lead to too many ($f(5,\frac{1}{2}) = 5$) spin particle permutations. Due to the closed shell $j\overline{j}$ being only moved to the new closed shell $v\overline{v}$, it is not increasing the number of open shells and therefore not increasing the spin degeneracy at all. Only the number of closed shells actually opened (additionally to the already present open shells) needs to be represented in the primitive spin function. In this example this leads to the primitive spin functions (above the $S_z = 0$ reference axis)
\begin{align*}
\theta_1 &= \alpha\beta\alpha \\
\theta_2 &= \alpha\alpha\beta\,, 
\end{align*}
composed of $\frac{\Delta O}{2} = 1$ $\alpha\beta$ pair as well as one additional $\alpha$ (to account for $S=\frac{1}{2}$). How to map the specific spin particles to spatial orbitals will be discussed in the next subsection. Here we shall only focus on the generation of primitive spin functions. 

\begin{center}
\begin{figure}
\includegraphics[scale=1]{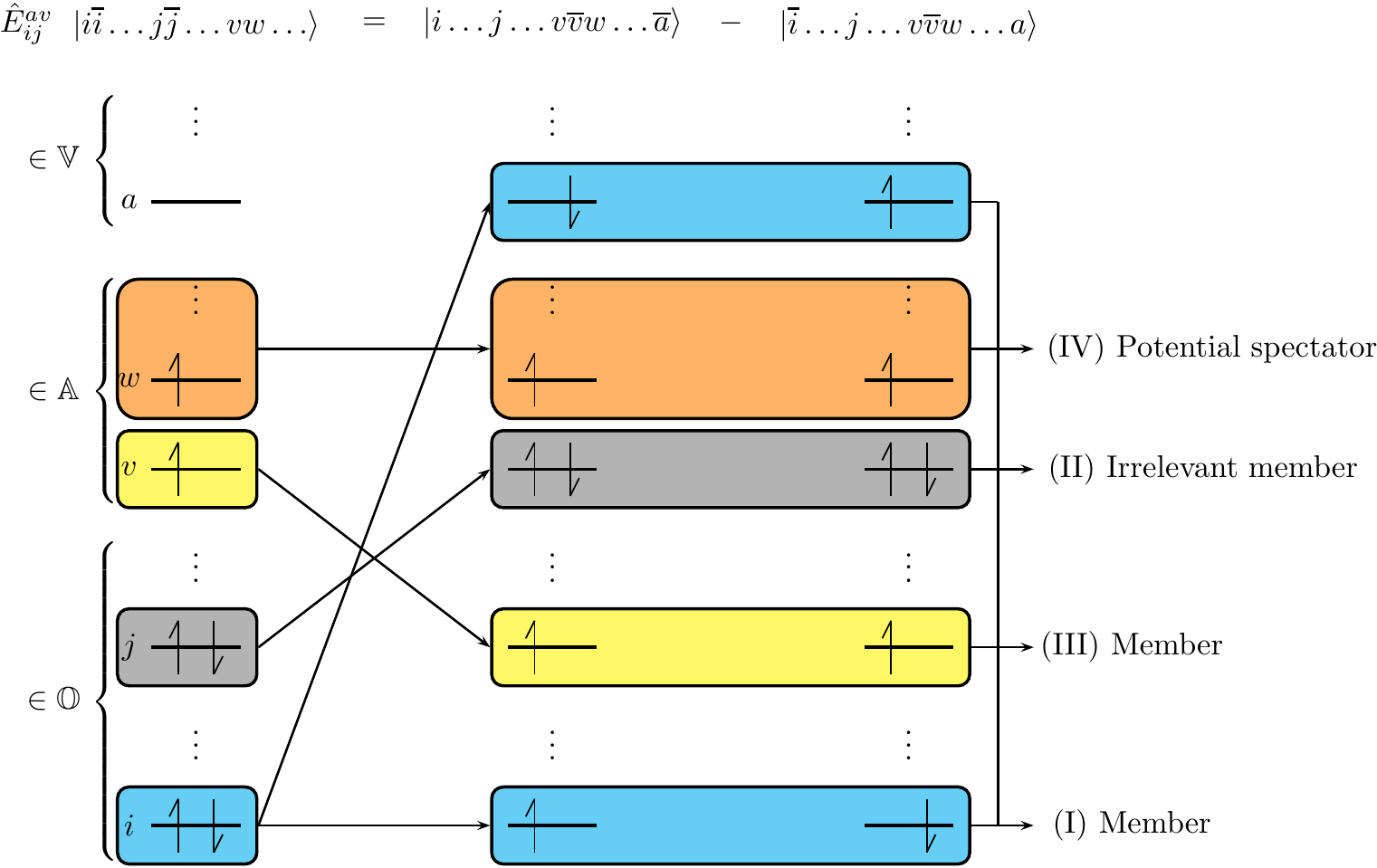}
\caption{\label{MemberSpectator.fig}Spatial substitution $\hat{E}_{ij}^{av}$ applied to an examplatory reference determinant depicted by orbital occupation schemes. The labels (I) to (IV) are used to distinguish between different particle substitution paths.}
\end{figure}
\end{center}

For a given initial spin function $\theta_1$, the problem of finding all primitive spin functions, that lead to a set of linearly independent spin eigenfunctions when projected, breaks down to finding all primitive spin functions that 
\begin{itemize}
\item[(i)] have the same number of $\alpha$'s and $\beta$'s as $\theta_1$ to conserve $S_z$ and 
\item[(ii)] always contain at least $n$ $\alpha$'s before a block of $n$ $\beta$'s to ensure that all path diagrams are completely above the $S_z=0$ reference axis.  
\end{itemize}
A simple algorithm incorporating (i) and (ii) works fine with arbitrary numbers of active electrons $2S$. Every single one of them participates one $\alpha$ electron at the end of $\theta_1$, which may be interchanged with all $\beta$ electrons. Therefore, this procedure will lead to specific permutations incorporating all $2S$ active indices explicitly. The generated $\hat{E}$ operators built using these permutations are only valid for up to $2S$ active indices. Any calculation involving higher quantum numbers will need additional operator generation. To achieve operators valid for arbitrary spin quantum numbers, we followed an improved generic approach:

A generic algorithm for arbitrary spin quantum numbers $S$ was developed where only the minimal number of non-redundant active indices per operator prototype is explicitly accounted for. Given the same example of the operator prototype $\hat{E}_{ij}^{av}$, this technique is motivated:

Consider the application of $\hat{E}_{ij}^{av}$ to an examplatory reference determinant as depicted in Figure \ref{MemberSpectator.fig}. Depending on the substitution path of different particles, we may distinguish between: 

\onecolumngrid
\twocolumngrid

\begin{itemize}
\item[(I)] Doubly occupied spatial orbitals opened through the substitution into an empty spatial orbital. 
\item[(II)] Doubly occupied spatial orbitals remaining doubly occupied or being shifted to another spatial orbital.
\item[(III)] Singly occupied spatial orbitals (part of the active space $\mathbb{A}$) moved by the applied spatial substitution operator.
\item[(IV)] Singly occupied spatial orbitals (part of the active space $\mathbb{A}$) remaining untouched by the applied spatial substitution operator.  
\end{itemize}

In this work, the orbitals of paths (I), (II) and (III) are labeled \textit{members} since they are explicitly appearing in the applied prototype (here $\hat{E}_{ij}^{av}$). As mentioned before, only the number of open shells ($\Delta O + 2S$) determines the spin degeneracy. Therefore, path (II) (not affecting $\Delta O$ nor $2S$) is irrelevant for the spin degeneracy and may be neglected in the primitive spin function generation for L\"owdin's method. Path (IV) is composed of unchanged singly occupied spatial orbitals, which are therefore labeled \textit{potential spectators}. While having no impact on $\Delta O$, an increased number of potential spectators leads to an increased spin quantum number and must therefore be taken into account.

To gain operator sets, which are correct for arbitrary spin quantum numbers, the minimal required amount of potential spectators per operator prototype needs to be determined.  
In Table \ref{Summationisbetter.tab}, all primitive spin functions for an increasing amount of potential spectators for the operator prototype $\hat{E}_{ij}^{av}$ are shown together with the respective $\hat{E}$ operators these functions would lead to. All primitive spin functions are split into 
\begin{itemize}
\item[(i)] its member part containing the ($\frac{\Delta O}{2} = 1$) $\alpha\beta$ pair for substitution path (I) as well as one single $\alpha$ particle for path (III) and 
\item[(ii)] its increasing potential spectator particles $\alpha$ resembling path (IV).
\end{itemize}

Please note that the inverse spatial orbital permutations were applied in the annihilator space $(ijvw\ldots)$ possibly breaking the annihilator ordering of the prototype. 

\begin{table}[h]
\caption{\label{Summationisbetter.tab}Primitive spin functions for an increasing amount of potential spectators (i.e. an increasing spin quantum number $S$) for the generation of linearly independent index permutations of the operator prototype $\hat{E}_{ij}^{av}$.}
\begin{threeparttable}
\centering
\begin{tabular}{c|c|ccc|cc}
& & \multicolumn{3}{c|}{\textbf{Spin Function}} & & \\
$S$& \textbf{Reference} & \multicolumn{2}{c}{\textbf{Member}} & \textbf{Spectator} & \multicolumn{2}{c}{\textbf{Operator}} \\ \hline
& & (I) & (III) & (IV) & &\\ \hline
\multirow{2}{*}{$\frac{1}{2}^{\text{a}}$} & \multirow{2}{*}{$\ket{i\overline{i}j\overline{j}v}$} 
& $\alpha\beta$ & $\alpha$ &  & $\hat{E}_{ij}^{av}$ \\ 
& & $\alpha\alpha$ & $\beta$ &  & $\hat{E}_{ji}^{av}$ \\ \hline
\multirow{3}{*}{$1$} & \multirow{3}{*}{$\ket{i\overline{i}j\overline{j}vw}$} 
& $\alpha\beta$ & $\alpha$ & $\alpha$ & $\hat{E}_{ij}^{av}$ \\
& & $\alpha\alpha$ & $\beta$ & $\alpha$ & $\hat{E}_{ji}^{av}$ \\
& & $\alpha\alpha$ & $\alpha$ & $\beta$ & $\hat{E}_{wji}^{avw}$ \\ \hline
\multirow{4}{*}{$\frac{3}{2}$} & \multirow{4}{*}{$\ket{i\overline{i}j\overline{j}vwx}$} 
& $\alpha\beta$ & $\alpha$ & $\alpha\alpha$ & $\hat{E}_{ij}^{av}$ \\
& & $\alpha\alpha$ & $\beta$ & $\alpha\alpha$ & $\hat{E}_{ji}^{av}$ \\
& & $\alpha\alpha$ & $\alpha$ & $\beta\alpha$ & $\hat{E}_{wji}^{avw}$ & \multirow{2}{*}{$\left.\rule{0pt}{12pt}\right\} \forall_{w\neq v\in\mathbb{A}} \hat{E}_{wji}^{avw}$}\\
& & $\alpha\alpha$ & $\alpha$ & $\alpha\beta$ & $\hat{E}_{xji}^{avx}$ \\ \hline
\multirow{5}{*}{$2$} & \multirow{5}{*}{$\ket{i\overline{i}j\overline{j}vwxy}$} 
& $\alpha\beta$ & $\alpha$ & $\alpha\alpha\alpha$ & $\hat{E}_{ij}^{av}$ \\
& & $\alpha\alpha$ & $\beta$ & $\alpha\alpha\alpha$ & $\hat{E}_{ji}^{av}$ \\
& & $\alpha\alpha$ & $\alpha$ & $\beta\alpha\alpha$ & $\hat{E}_{wji}^{avw}$ & \multirow{3}{*}{$\left.\rule{0pt}{18pt}\right\} \forall_{w\neq v\in\mathbb{A}} \hat{E}_{wji}^{avw}$}\\
& & $\alpha\alpha$ & $\alpha$ & $\alpha\beta\alpha$ & $\hat{E}_{xji}^{avx}$ \\ 
& & $\alpha\alpha$ & $\alpha$ & $\alpha\alpha\beta$ & $\hat{E}_{yji}^{avy}$
\end{tabular}
\begin{tablenotes}
\item[a] The fragmentation into members (I) and (III) as well as potential spectators (IV) for the doublet case is explicitly shown in Figure \ref{MemberSpectator.fig}.
\end{tablenotes}
\end{threeparttable}
\end{table}  

With an increasing amount of potential spectators (resembling an increasing spin quantum number $S = S_z$), the number of $\alpha$ electrons in the reference as well as the primitive spin function increases. In the doublet case ($S = \frac{1}{2}$), there are no potential spectators and therefore only two possible spin functions leading to two operators are found. This picture changes in the triplet case ($S = 1$), where a new additional function ($\alpha\alpha\alpha\beta$) leads to the augmented operator $\hat{E}_{wji}^{avw}$ including the spectator index $w$. In the following quartet ($S = \frac{3}{2}$), quintet ($S = 2$), etc. cases, only repeated transpositions of the single $\beta$ electron with all potential $\alpha$ spectator electrons arise. These do not lead to fundamentally new operators. they only produce the same type of operator ($\hat{E}_{wji}^{avw}$) for the increasing active space $\mathbb{A}\setminus\{v\} = \{ w, x, y,\ldots\}$. To fully determine all possible primitive spin functions (for arbitrary $S$) it is therefore sufficient to consider $\frac{\Delta O}{2}$ potential spectators only. Furthermore, only primitive spin functions with differing member parts need to be taken into account. All functions containing identical member parts can be taken care of by an index iteration in the active space. Since this iteration requires special care wrt. active indices entering the prototypes from different augmentations, it will be discussed seperately in subsection \ref{IndexSymmetry}. 

To further clarify the primitive spin function generation, consider the following examples explicitly showing all primitive spin functions (required and omitted) for different operator prototypes:

\newpage
\begin{itemize}
\item[(1)] $\hat{E}_{ijjk}^{abcc}$ of type $\Delta O = 4$ 
\begin{center}
\begin{tabular}{cccccc}
\multicolumn{2}{c}{\textbf{Function}} & & \textbf{Permutation} & & \textbf{Operator}\\ \hline
$\alpha\beta\alpha\beta$ & $\alpha\alpha$ & $\longrightarrow$ & $()$ & $\longrightarrow$ & $\hat{E}_{ijjk}^{abcc}$ \\
$\alpha\beta\alpha\alpha$ & $\beta\alpha$ & $\longrightarrow$ & $(j,v)$ & $\longrightarrow$ & $\hat{E}_{ivjkj}^{abccv}$  \\
$\alpha\beta\alpha\alpha$ & $\alpha\beta$ &  &  &  &   \\[-1.7ex]
\hline\noalign{\vspace{\dimexpr 1.7ex-\doublerulesep}}
$\alpha\alpha\beta\beta$ & $\alpha\alpha$ & $\longrightarrow$ & $(i,j)$ & $\longrightarrow$ & $\hat{E}_{jijk}^{abcc}$ \\
$\alpha\alpha\beta\alpha$ & $\beta\alpha$ & $\longrightarrow$ & $(j,v)(i,j)$ & $\longrightarrow$ & $\hat{E}_{vijkj}^{abccv}$ \\
$\alpha\alpha\beta\alpha$ & $\alpha\beta$ &  &  &  &   \\[-1.7ex]
\hline\noalign{\vspace{\dimexpr 1.7ex-\doublerulesep}}
$\alpha\alpha\alpha\beta$ & $\beta\alpha$ & $\longrightarrow$ & $(i,v)$ & $\longrightarrow$ & $\hat{E}_{vjjki}^{abccv}$  \\
$\alpha\alpha\alpha\beta$ & $\alpha\beta$ &  &  &  &   \\[-1.7ex]
\hline\noalign{\vspace{\dimexpr 1.7ex-\doublerulesep}}
$\alpha\alpha\alpha\alpha$ & $\beta\beta$ & $\longrightarrow$ & $(i,v)(j,w)$ & $\longrightarrow$ & $\hat{E}_{vwjkij}^{abccvw}$  \\
\end{tabular}
\end{center}

Due to the the additionally created open shells $\Delta O = 4$, two potential spectators resembling indices $v$ and $w$ need to be introduced. Furthermore, the size of the member part is four. All primitive spin functions, which contain the exact same member part of another primitive spin function can be neglected (they need to be taken care of by further index iteration).

\item[(2)] $\hat{E}_{ijjkk}^{abccv}$ of type $\Delta O = 2$ 
\begin{center}
\begin{tabular}{cccccc}
\multicolumn{2}{c}{\textbf{Function}} & & \textbf{Permutation} & & \textbf{Operator}\\ \hline
$\alpha\beta\alpha$ & $\alpha$ & $\longrightarrow$ & $()$ & $\longrightarrow$ & $\hat{E}_{ijjkk}^{abccv}$ \\
$\alpha\alpha\beta$ & $\alpha$ & $\longrightarrow$ & $(i,j)$ & $\longrightarrow$ & $\hat{E}_{jijkk}^{abccv}$ \\
$\alpha\alpha\alpha$ & $\beta$ & $\longrightarrow$ & $(i,w)$ & $\longrightarrow$ & $\hat{E}_{wjjkki}^{abccvw}$
\end{tabular}
\end{center}

In this case, the operator creates two additional open shells. Due to the explicit occurrence of the active index $v$ in the prototype, the member part is composed of three particles. Furthermore, one potential spectator resembling index $w$ needs to be introduced. All three primitive spin functions have differing member parts and are therefore mandatory.  

\item[(3)] $\hat{E}_{ijv}^{abc}$ of type $\Delta O = 4$
\begin{center}
\begin{tabular}{cccccc}
\multicolumn{2}{c}{\textbf{Function}} & & \textbf{Permutation} & & \textbf{Operator}\\ \hline
$\alpha\beta\alpha\beta\alpha$ & $\alpha\alpha$ & $\longrightarrow$ & $()$ & $\longrightarrow$ & $\hat{E}_{ijv}^{abc}$ \\
$\alpha\beta\alpha\alpha\beta$ & $\alpha\alpha$ & $\longrightarrow$ & $(j, v)$ & $\longrightarrow$ & $\hat{E}_{ivj}^{abc}$ \\
$\alpha\beta\alpha\alpha\alpha$ & $\beta\alpha$ & $\longrightarrow$ & $(j, w)$ & $\longrightarrow$ & $\hat{E}_{iwvj}^{abcw}$ \\
$\alpha\beta\alpha\alpha\alpha$ & $\alpha\beta$ &  &  &  &   \\[-1.7ex]
\hline\noalign{\vspace{\dimexpr 1.7ex-\doublerulesep}}
$\alpha\alpha\beta\beta\alpha$ & $\alpha\alpha$ & $\longrightarrow$ & $(i, j)$ & $\longrightarrow$ & $\hat{E}_{jiv}^{abc}$ \\
$\alpha\alpha\beta\alpha\beta$ & $\alpha\alpha$ & $\longrightarrow$ & $(j, v)(i, j)$ & $\longrightarrow$ & $\hat{E}_{vij}^{abc}$ \\
$\alpha\alpha\beta\alpha\alpha$ & $\beta\alpha$ & $\longrightarrow$ & $(j, w)(i, j)$ & $\longrightarrow$ & $\hat{E}_{wivj}^{abcw}$ \\
$\alpha\alpha\beta\alpha\alpha$ & $\alpha\beta$ &  &  &  &   \\[-1.7ex]
\hline\noalign{\vspace{\dimexpr 1.7ex-\doublerulesep}}
$\alpha\alpha\alpha\beta\beta$ & $\alpha\alpha$ & $\longrightarrow$ & $(i, v)$ & $\longrightarrow$ & $\hat{E}_{vji}^{abc}$ \\
$\alpha\alpha\alpha\beta\alpha$ & $\beta\alpha$ & $\longrightarrow$ & $(i, w)$ & $\longrightarrow$ & $\hat{E}_{wjvi}^{abcw}$ \\
$\alpha\alpha\alpha\beta\alpha$ & $\alpha\beta$ &  &  &  &   \\[-1.7ex]
\hline\noalign{\vspace{\dimexpr 1.7ex-\doublerulesep}}
$\alpha\alpha\alpha\alpha\beta$ & $\beta\alpha$ & $\longrightarrow$ & $(i, v)(j, w)$ & $\longrightarrow$ & $\hat{E}_{vwij}^{abcw}$ \\
$\alpha\alpha\alpha\alpha\beta$ & $\alpha\beta$ &  &  &  &   \\[-1.7ex]
\hline\noalign{\vspace{\dimexpr 1.7ex-\doublerulesep}}
$\alpha\alpha\alpha\alpha\alpha$ & $\beta\beta$ & $\longrightarrow$ & $(i, w)(j, x)$ & $\longrightarrow$ & $\hat{E}_{wxvij}^{abcwx}$
\end{tabular}
\end{center}

In this last example, four additional open shells are created such that an augmentation by two potential spectators ($w$ and $x$) is necessary. There is one explicitly occurring active index $v$ leading to a member part of size five. Again, all primitive spin functions, which contain the exact same member part of another primitive spin function can be neglected.
\end{itemize}

\subsection{Spin Particle $\longrightarrow$ Spatial Orbital Mapping}
\label{SpinSpatialMapping}

One of the central parts of this work is the conversion from spin particle permutations (occurring from L\"owdin's method c.f. section \ref{LoewdinsMethod}) to spatial orbital permutations via the spatial high spin reference configuration. As stated before, it is possible to do the conversion by solving a topological problem. To tackle this problem, specific diagrams, which we will call substitution path diagrams, are useful.

For arbitrary spin orbital $\hat{X}$ or spatial orbital $\hat{E}$ substitution operators, it is possible to define a substitution path diagram (SPD), where annihilated indices are connected to created indices from below by single arrows facing upwards. In case of the usual single reference spin orbital CC, all occurring substitution operators may be cast into the form 
\begin{equation*}
\hat{X}_{\tilde{i} < \tilde{j} < \tilde{k} < \ldots}^{\tilde{a} < \tilde{b} < \tilde{c} < \ldots}\,,
\end{equation*}
where a tilde shall denote spin orbital indices. Therefore, the corresponding SPDs always take the form 
\begin{center}
\includegraphics[scale=1]{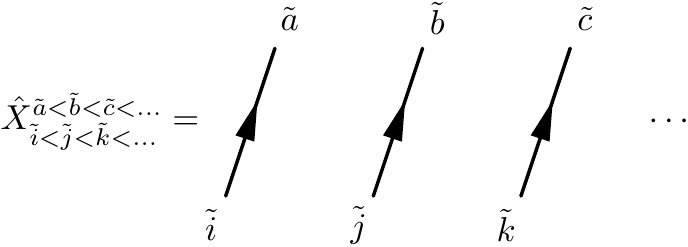}\,.
\end{center}

For any $\nu$-fold substitution, there are $\nu$ disconnected diagram fragments consisting of single lines. This trivial picture changes, if spin-adapted prototypes $\hat{E}$ are considered. Arbitrary $\hat{E}$ prototypes (according to the routine described in subsection \ref{PrototypeGeneration}) possess the form 
\begin{equation*}
\hat{E}_{p_1 \le \ldots \le p_{\nu}}^{q_1 \ldots q_{\nu}}\,,
\end{equation*}
where index notations according to subsection \ref{HighSpinOpenShellReference} were used. Due to the smaller equal relation in the annihilator space and the absence of any relation in the creator space, the spatial indices $p_1\ldots p_{\nu}$ and $q_1\ldots q_{\nu}$ may have multiple occurrences in the corresponding SPDs. All $\hat{E}$ prototype SPDs can easily be assigned to the number of open shells $\Delta O$ additionally created by the latter operator (when applied to the appropriate high spin reference). In Table \ref{OpenShellsToSPD.tab}, a few example SPDs sorted by $\Delta O$ are shown. 

\begin{table}[h]
\caption{\label{OpenShellsToSPD.tab}Exemplary substitution path diagrams (SPDs) for different $\hat{E}$ prototypes sorted by the number of open shells $\Delta O$ created when applied to the appropriate reference CSF.}
\begin{tabular}{c|Sc}
$\Delta O$ & Substitution path diagrams \\ \hline 
0 & \cincludegraphics[scale=1]{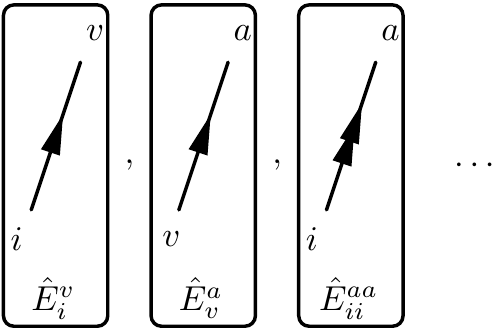} \\ \hline
2 & \cincludegraphics[scale=1]{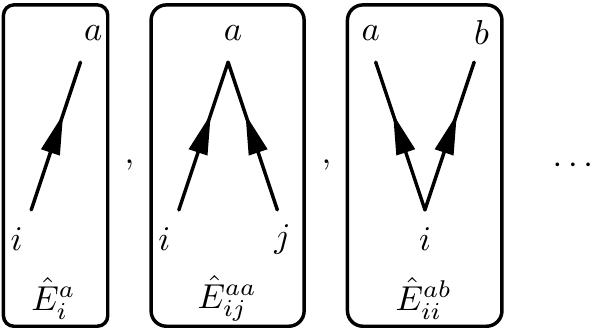} \\ \hline 
4 & \cincludegraphics[scale=1]{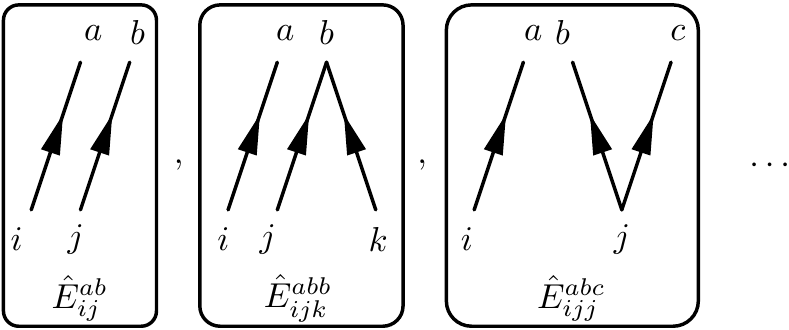}
\end{tabular}
\end{table}

The almost trivial assignment to the corresponding $\Delta O$ level is due to the fact that each SPD always consists of $\frac{\Delta O}{2}$ disconnected fragments, which create additional open shells. As stated in the last subsection \ref{IndexPermutation}, primitive spin functions from the application of L\"owdin's method need to consist of $\frac{\Delta O}{2}$ pairs of $\alpha$ and $\beta$ electrons giving rise to an anchor point to map spin particle permutations to spatial orbital permutations. 

Before several examples of this procedure are presented, we would like to show that arbitrary SPDs contain exactly $\frac{\Delta O}{2}$ disconnected fragments, which create additional open shells:

The number of additional open shells $\Delta O$ generated by a specific operator $\hat{E}_{p_1\ldots p_{\nu}}^{q_1\ldots q_{\nu}}$ is given by 
\begin{equation*}
\Delta O = O\left(\hat{E}_{p_1\ldots p_{\nu}}^{q_1\ldots q_{\nu}}\ket{\Psi_0}\right) - O\left(\ket{\Psi_0}\right)\, ,
\end{equation*}
where $O(x)$ shall denote the absolute number of open shells of CSF $x$ and $\ket{\Psi_0}$ shall be an appropriate reference CSF.
For any operator $\hat{E}_{p_1\ldots p_{\nu}}^{q_1\ldots q_{\nu}}$, it is
\begin{equation*}
\hat{E}_{p_1\ldots p_{\nu}}^{q_1\ldots q_{\nu}} = \prod_i \hat{E}_{\{p\}_i}^{\{q\}_i}
\end{equation*}
where $\{p\}_i$ and 
$\{q\}_i$ denote subsets of $\{p_1\ldots p_{\nu}\}$ and 

\onecolumngrid 
\begin{center}
\begin{figure}[h]
\includegraphics[scale=1]{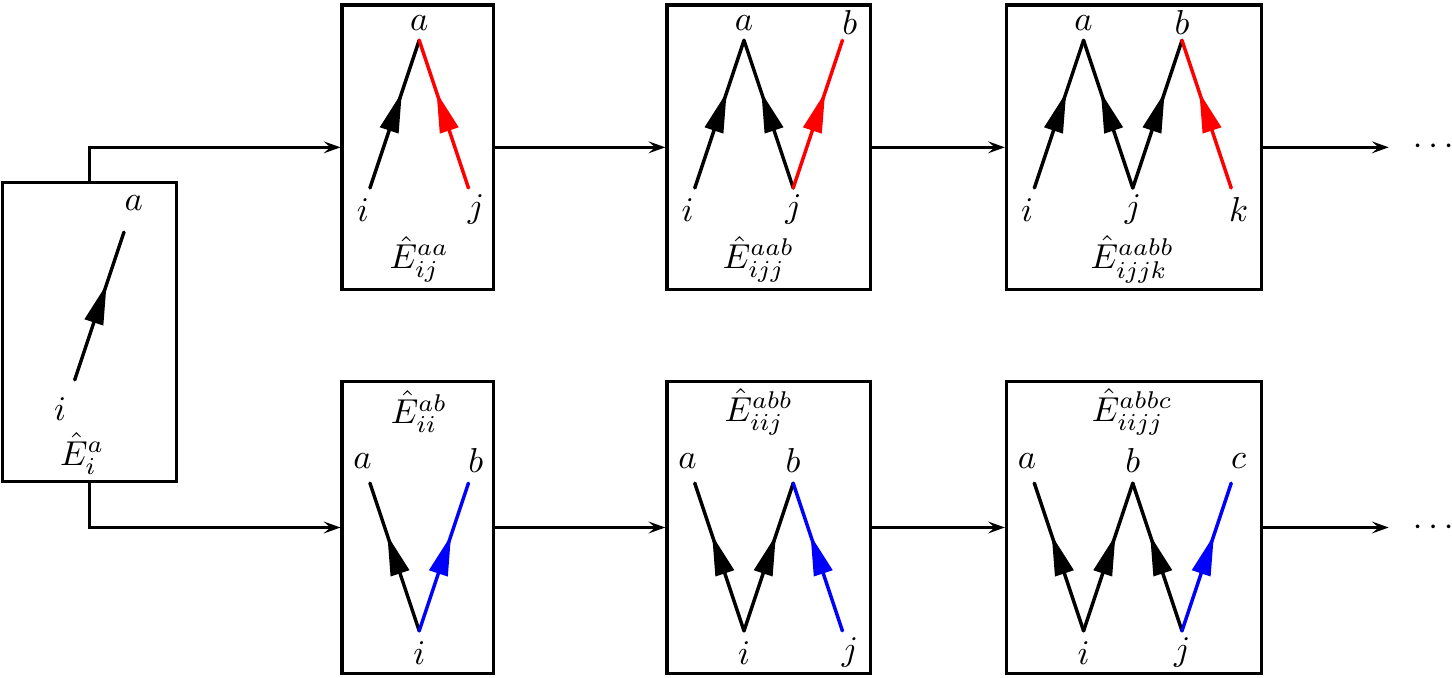}
\caption{\label{Aufbau.fig} Augmentation (\glqq{s}naking\grqq{) of the exemplary SPD of $\hat{E}_i^a$ starting with occupied indices (red) and virtual indices (blue).} }
\end{figure}
\end{center}
\twocolumngrid

$\{q_1\ldots q_{\nu}\}$, respectively with
\begin{equation*} 
\left\{ \{p\}_i \cup \{q\}_i \right\} \cap \left\{ \{p\}_j \cup \{q\}_j \right\} = \emptyset \quad\forall_{i\neq j}\,.
\end{equation*}
All of these fragments possess individual disconnected SPDs. Furthermore, $\Delta O$ is additive with 
\begin{equation}
\label{ProofSum.eq}
\Delta O = \sum_i \Delta o_i
\end{equation}
for 
\begin{equation*}
\Delta o_i = O\left(\hat{E}_{\{p\}_i}^{\{q\}_i}\ket{\Psi_0}\right) - O\left(\ket{\Psi_0}\right)\, .
\end{equation*}

In general it is 
\begin{equation}
\label{Proof02.eq}
\Delta o_i \in \{0, 2\} \quad\forall_i 
\end{equation}
since 
\begin{itemize}
\item[(i)] the particle number is conserved such that $\Delta o_i$ must be even,
\item[(ii)] $\ket{\Psi_0}$ is a high spin reference such that $\Delta o_i \ge 0$ and 
\item[(iii)] no single fragment of type $\Delta o_i > 2$ is possible. 
\end{itemize}

While (i) and (ii) are trivial, (iii) needs to be shown in greater detail. Consider the simplest $\Delta o_i = 2$ fragment $\hat{E}_i^a$ as visualized in Figure \ref{Aufbau.fig} on the left. If a single fragment of type $\Delta o_i \ge 4$ was possible, the fragment $\hat{E}_i^a$ must be augmentable in some fashion to reach higher $\Delta o_i$ values while keeping its connectivity intact. Any connected augmentation of $\hat{E}_i^a$ is possible via (1) doubling the annihilator $i$ or (2) doubling the creator $a$. Depending on the initially doubled index, two augmentation routes (visualized in Figure \ref{Aufbau.fig}) are possible. Please note that augmentations to active indices are not considered, since they would decrease the $\Delta o_i$ level to $0$ as in e.g. $\hat{E}_{ii}^{av}$ or $\hat{E}_{iv}^{aa}$.
 There can be no more than two incoming or outgoing lines per index since every spatial orbital can contain two electrons at most. In Figure \ref{Aufbau.fig}, both augmentation routes show a \glqq{s}naking\grqq{} behavior in which the total number of additionally created open shells stays constant at 2 (countable through the number of single line endings). Therefore, an augmentation to a single fragment of $\Delta o_i > 2$ is impossible.  

Due to (\ref{ProofSum.eq}) and (\ref{Proof02.eq}), it is 
\begin{align*}
\frac{\Delta O}{2} &= \sum_i \frac{\Delta o_i}{2} = \sum_i \frac{2n_i}{2} \text{ with } n_i \in \{0,1\} \\
&= \sum_i n_i = \#(n_i = 1 \quad \forall_i)\,, 
\end{align*}
where $\#(n_i = 1 \quad \forall_i)$ denotes the amount of $\Delta o_i = 2$ fragments in the operator. This completes the proof that 
each $\hat{E}$ operator must contain exactly $\frac{\Delta O}{2}$ disconnected SPD fragments $i$ of type $\Delta o_i = 2$.

To map these fragments to spin particle permutations, all fragments $i$ of type $\Delta o_i = 2$ are assigned to inidividual $\alpha\beta$ pairs. The remaining fragments $j$ of type $\Delta o_j = 0$ are (i) not mapped at all or (ii) mapped to single $\alpha$ particles if they contain an acitve index. 
Hereby, the assignment of single fragments to $\alpha\beta$ pairs or single $\alpha$ particles is arbitrary as long as each fragment maps to a different pair or particle. Spin particle permutations can now be translated to spatial orbital permutations using anyone of the occurring spatial indices from the respective fragments.

To further clarify the procedure presented in this subsection, consider the following examples, where the primitive reference spin functions (resembling the examples of subsection \ref{IndexPermutation}) are explicitly mapped to the corresponding SPDs. Please note that examples (2) and (3) are specially crafted to represent educational illustrations. With the established rules of subsection \ref{IndexPermutation} both examples (2) and (3) would require one additional spectator each. 

\begin{itemize}
\item[(1)] $\hat{E}_{ijjk}^{abcc}$ for a $S = S_z = 1$ reference $\ket{i\overline{i}j\overline{j}k\overline{k}\ldots vw}$.
\begin{center}
\includegraphics[scale=1]{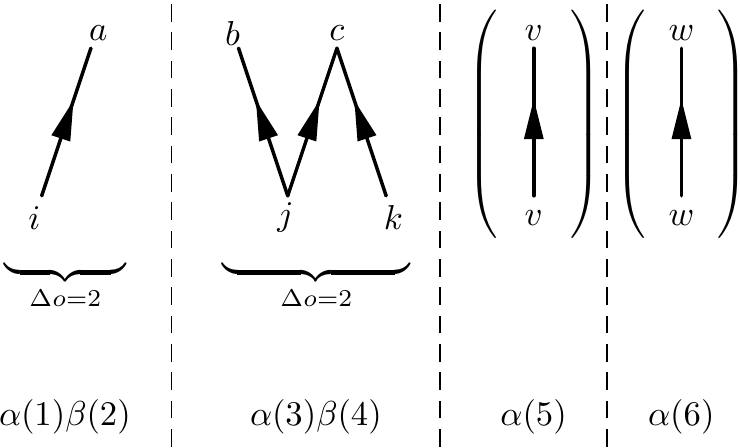}
\end{center}
The two SPD fragments are of type $\Delta o = 2$ and are mapped to different $\alpha\beta$ pairs. Any transposition involving e.g. $\alpha(3)$ or $\beta(4)$ can now be translated to a transposition of spatial orbital indices $j$ or $k$ in the annihilator space or spatial orbital indices $b$ or $c$ in the creator space. Due to the triplet reference, two temporary spectator substitutions need to be introduced and mapped to single $\alpha$ electrons.
\item[(2)] $\hat{E}_{ijjkk}^{abccv}$ for a $S = S_z = \frac{1}{2}$ reference $\ket{i\overline{i}j\overline{j}k\overline{k}\ldots v}$. 
\begin{center}
\includegraphics[scale=1]{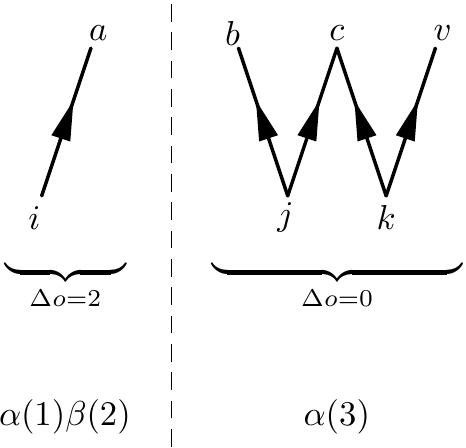}
\end{center}
The left SPD fragment of type $\Delta o = 2$ is mapped to the only $\alpha\beta$ pair, while the right $\Delta o = 0$ fragment is mapped to the single $\alpha(3)$ electron since the active index $v$ is connected to this fragment. Due to the spin state of $S = S_z = \frac{1}{2}$ of the reference CSF, there is no further temporary active index (single $\alpha$ electron) mapped. 
\newpage
\item[(3)] $\hat{E}_{ijv}^{abc}$ for a $S = S_z = 1$ reference $\ket{i\overline{i}j\overline{j}\ldots vw}$.
\begin{center}
\includegraphics[scale=1]{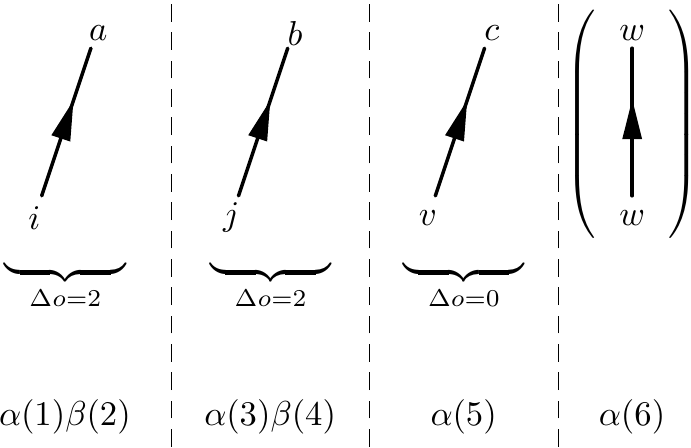}
\end{center}
In this example, two SPD fragments of type $\Delta o = 2$ are mapped to different $\alpha\beta$ pairs. The $\Delta o = 0$ fragment is connected with the active index $v$ and therefore mapped to $\alpha(5)$. Due to the spin quantum number $S = S_z = 1$, one more temporary $\mathbb{A}\rightarrow\mathbb{A}$ substitution is necessary.  
\end{itemize}

\subsection{Canonical Index Ordering}
\label{IndexSymmetry}

The last step in the generation of linearly independent $\hat{E}$ operators is the application of a proper index ordering for occupied ($\mathbb{O}$), virtual ($\mathbb{V}$) and active ($\mathbb{A}$) space indices. By construction (c.f. subsection \ref{PrototypeGeneration}), the occupied and virtual indices $i,j,k,\ldots$ and $a,b,c,\ldots$, respectively, were fixed with 
\begin{align}
\label{OccRelation.eq}
i < j &< k < \ldots \\
a < b &< c < \ldots\,. 
\label{VirRelation.eq}
\end{align}

Only the relation of all active indices $v, w, x, \ldots$, which may occur in either creator, annihilator or both index spaces, remains to be determined. In general, there are three ways active indices may enter specific $\hat{E}$ operators: 
\begin{itemize}
\item[(i)] by prototype augmentation $\mathbb{O}\rightarrow\mathbb{A}$ in the creator space (e.g. $\hat{E}_{i}^{a} \rightarrow \hat{E}_{ij}^{av}$),
\item[(ii)] by prototype augmentation $\mathbb{A}\rightarrow\mathbb{V}$ in the annihilator space (e.g. $\hat{E}_{i}^{a}\rightarrow \hat{E}_{iv}^{ab}$)
\item[(iii)] or by index permutation using temporary $\mathbb{A}\rightarrow\mathbb{A}$ spectator substitutions in both the creator and the annihilator space (e.g. $(i, v)\hat{E}_{iv}^{av} = \hat{E}_{vi}^{av}$).
\end{itemize}

The relation of all active indices within cases (i), (ii) or (iii) can be fixed by construction to yield 
\begin{subequations}
\label{ActRelation1.eq}
\begin{align}
\label{ActiveRestriction1a.eq}
v^{\text{(i)}\phantom{ii}} &< w^{\text{(i)}\phantom{ii}} < x^{\text{(i)}\phantom{ii}} < \ldots \\
\label{ActiveRestriction1b.eq}
v^{\text{(ii)}\phantom{i}} &< w^{\text{(ii)}\phantom{i}} < x^{\text{(ii)}\phantom{i}} < \ldots \\
\label{ActiveRestriction1c.eq}
v^{\text{(iii)}} &< w^{\text{(iii)}} < x^{\text{(iii)}} < \ldots\,.
\end{align}
\end{subequations}

We can imply that no common indices occur in different sets $\mathbb{A}^{\text{(i)}}$, $\mathbb{A}^{\text{(ii)}}$ and $\mathbb{A}^{\text{(iii)}}$ such that no unintentional spectators are formed:
\begin{equation}
\label{ActRelation2.eq}
\mathbb{A}^{\text{(i)}} \cap \mathbb{A}^{\text{(ii)}} = \mathbb{A}^{\text{(i)}} \cap \mathbb{A}^{\text{(iii)}} = \mathbb{A}^{\text{(ii)}} \cap \mathbb{A}^{\text{(iii)}} = \emptyset
\end{equation}

The combined relations \eqref{OccRelation.eq}, \eqref{VirRelation.eq}, \eqref{ActRelation1.eq} and \eqref{ActRelation2.eq} need to be applied to all generated $\hat{E}$ operators to yield the final set $\mathbb{E}$ of linearly independent operators. To further clarify this procedure, consider the following examples, which now reside on the results from examples (1), (2) and (3) of the previous subsections \ref{IndexPermutation} and \ref{SpinSpatialMapping}. 

\begin{itemize}
\item[(1)] $\hat{E}_{ijjk}^{abcc}$ of type $\Delta O = 4$ 
\begin{align*}
\mathbb{E}_{ijjk}^{abcc} = &\left\{\rule{0pt}{15pt}\;\left\{\hat{E}_{ijjk}^{abcc} ; \hat{E}_{jijk}^{abcc}\right\}\right. \cup  \\
&\hspace{12pt}\left\{\hat{E}_{ivjkj}^{abccv} ; \hat{E}_{vijkj}^{abccv} ; \hat{E}_{vjjki}^{abccv} \rule{0pt}{10pt}\middle\vert\, v\in\mathbb{A}\right\} \cup \\
&\left.\hspace{12pt}\left\{\hat{E}_{vwjkij}^{abccvw}\,\middle\vert\, (v<w) \in\mathbb{A}\right\}\rule{0pt}{15pt}\right\}
\end{align*}

Due to the abundance of active indices in the original prototype, all active indices in permuted operators originate from spectator substitutions (type (iii)). Therefore, they either show the relation $v\in\mathbb{A}$ for one spectator or $(v < w)\in\mathbb{A}$ for two spectators.
 
\item[(2)] $\hat{E}_{ijjkk}^{abccv}$ of type $\Delta O = 2$ 
\begin{align*}
\mathbb{E}_{ijjkk}^{abccv} = &
\left\{
  \rule{0pt}{15pt}\; 
  \left\{
    \hat{E}_{ijjkk}^{abccv} ; \hat{E}_{jijkk}^{abccv} \,\middle\vert\, v\in\mathbb{A} 
  \right\} 
  \cup\right. \\
&\hspace{12pt}
\left.
  \left\{
    \hat{E}_{wjjkki}^{abccvw} 
    \rule{0pt}{10pt}\middle\vert\, (v\neq w)\in\mathbb{A}
  \right\} 
  \rule{0pt}{15pt}
\right\} 
\end{align*}

In this case, there are active indices from different origins. The index $v$ is already present in the original prototype (type (i)), while the index $w$ is a spectating index (type (iii)). The permuted operator $\hat{E}_{wjjkki}^{abccvw}$ therefore needs to be built for all $(v \neq w)\in\mathbb{A}$. 

\item[(3)] $\hat{E}_{ijv}^{abc}$ of type $\Delta O = 4$ 
\begin{align*}
\mathbb{E}_{ijv}^{abc} = &
\left\{
  \rule{0pt}{15pt}\; 
  \left\{
    \hat{E}_{ijv}^{abc} ; \hat{E}_{ivj}^{abc} ; \hat{E}_{jiv}^{abc}; \hat{E}_{vij}^{abc}; \hat{E}_{vji}^{abc} \,\middle\vert\, v\in\mathbb{A}
  \right\} \cup
\right.\\
&\hspace{12pt}
\left\{
  \hat{E}_{iwvj}^{abcw} ; \hat{E}_{wivj}^{abcw} ; \hat{E}_{wjvi}^{abcw} ; \hat{E}_{vwij}^{abcw} \,\middle\vert\, (v\neq w)\in\mathbb{A}
\right\} 
\cup \\
&\left.\hspace{12pt}
\left\{
  \hat{E}_{wxvij}^{abcwx} \,\middle\vert\, (v \neq (w < x)) \in\mathbb{A}
\right\} 
\;\rule{0pt}{15pt}
\right\} 
\end{align*}

This last example has the active indices $v$ originating from type (ii) as well as the indices $w$ and $x$ originating from type (iii). Operators involving no spectator must therefore be built for all $v\in\mathbb{A}$, operators involving one spectator $w$ for all $(v\neq w)\in\mathbb{A}$ and operators involving two spectators $w$ and $x$ for all $(v\neq(w < x))\in\mathbb{A}$.

\end{itemize}

\section{Application}
\label{Application}

\subsection{Linear Independence and Completeness Check}

To check if the presented scheme to generate spatial substitution operators indeed produces linearly independent and spin-complete operators, we confirmed 
\begin{itemize}
\item[(I)] their dimensionality (the number of generated operators) and 
\item[(II)] their linear independence. 
\end{itemize}

To check for the correct dimensionality (I), the number of generated operators was compared to the dimensionality of the full CSF-space $d(n,S,b)$ for $n$ electrons with spin quantum number $S$ in $b$ spatial orbitals, which is given by the Weyl-Robinson-dimension formula (see e.g. \cite{Pauncz1979}): 
\begin{equation}
\label{RWeyl.eq}
d(n,S,b) = \frac{2S+1}{b+1}\binom{b+1}{\frac{1}{2}n+S+1}\binom{b+1}{\frac{1}{2}n-S}
\end{equation}
The check their linear independence (II), the generated operators were sorted into sets $\{\hat{E}\}_{\Phi}$ such that operators leading to the same spatial function $\Phi$ when applied to the reference CSF are in the
same set. Every set of operators was applied to the reference CSF and the resulting CSFs where represented in a minimal determinant basis. The representations were gathered as row vectors in a matrix $\uuline{C}$, which was decomposed in a rank revealing Householder-QR-decomposition. If the ranks of all matrices $\uuline{C}$ for all spatial functions $\Phi$ are exactly equal to the number of their row vectors, the set of all generated $\hat{E}$ operators leads to linearly independent CSFs when applied to the reference. 

In Table \ref{RWeyl.tab}, the results for spatial substitution operators for all high spin reference CSFs composed of $n=2$ to $n=10$ particles in all possible high spin states ($S=0$ to $S=5$) in $b=3(n_o + n_a)$ spatial orbitals of up to spatial substitution rank $n$ are summarized.

\begin{table}
\caption{\label{RWeyl.tab} The number of generated operators compared to the dimensionality of the full CSF basis for every $(n,S,b)$-tuple considered in this work.}
\begin{threeparttable}
\centering
\begin{tabular}{ccc|rr}
$n$ & $S$ & $b$ & \#Operators$^\text{a}$ & $d(n,S,b)^\text{b}$ \\ \hline
2 & 0     & 3  & 5    & 6     \\
2 & 1     & 6  & 14   & 15    \\
3 & $1/2$ & 6  & 69   & 70    \\
3 & $3/2$ & 9  & 83   & 84    \\
4 & 0     & 6  & 104  & 105   \\
4 & 1     & 9  & 629  & 630   \\
4 & 2     & 12 & 494  & 495   \\
5 & $1/2$ & 9  & 1,889 & 1,890  \\ 
5 & $3/2$ & 12 & 5,147 & 5,148  \\
5 & $5/2$ & 15 & 3,002 & 3,003  \\
6 & 0     & 9  & 2,519    & 2,520     \\
6 & 1     & 12 & 23,165   & 23,166    \\
6 & 2     & 15 & 40,039   & 40,040    \\
6 & 3     & 18 & 18,563   & 18,564    \\
7 & $1/2$ & 12 & 56,627   & 56,628    \\
7 & $3/2$ & 15 & 240,239  & 240,240   \\
7 & $5/2$ & 18 & 302,327  & 302,328   \\
7 & $7/2$ & 21 & 116,279  & 116,280   \\
8 & 0     & 12 & 70,784   & 70,785    \\
8 & 1     & 15 & 840,839  & 840,840   \\
8 & 2     & 18 & 2,267,459 & 2,267,460  \\
8 & 3     & 21 & 2,238,389 & 2,238,390  \\
8 & 4     & 24 & 735,470  & 735,471   \\
9  & $1/2$ & 15 &   1,821,819  &   1,821,820  \\ 
9  & $3/2$ & 18 &  10,279,151  &  10,279,152  \\
9  & $5/2$ & 21 &  20,145,509  &  20,145,510  \\
9  & $7/2$ & 24 &  16,343,799  &  16,343,800  \\
9  & $9/2$ & 27 &   4,686,824  &   4,686,825  \\
10 & 0     & 15 &   2,186,183  &   2,186,184  \\
10 & 1     & 18 &  30,837,455  &  30,837,456  \\
10 & 2     & 21 & 111,919,499  & 111,919,500  \\
10 & 3     & 24 & 171,609,899  & 171,609,900  \\
10 & 4     & 27 & 118,107,989  & 118,107,990  \\ 
10 & 5     & 30 &  30,045,014  &  30,045,015  
\end{tabular}
\begin{tablenotes} 
\item[a] Generated (checked for linear independence)
\item[b] Formula \eqref{RWeyl.eq}
\end{tablenotes}
\end{threeparttable}
\end{table}

All operator sets were found to produce linearly independent CSFs and showed the correct dimensionality of $d(n,S,b)-1$ (one CSF being the reference with no corresponding substitution operator) proving their completeness. Therefore, we conclude that the proposed operator generation scheme works for up to 10-fold substitutions. 

\subsection{Proof of Concept Test Calculation}
\label{TestCalculation}

A proof of concept implementation for spin-adapted and spin-complete (SASC) CC utilizing the generated spatial substitutions in this work was developed. Following an ROHF calculation using the PySCF \cite{PYSCF, Sun_2015} program package, all operators (cluster and Hamiltonian) were represented in the FCI CSF basis. This allows the BCH-series to be evaluated using matrix commutators only. Therefore, any implementational difficulties arrising from the non-commutative cluster operators or an overlapping occupied and virtual space do not have to be dealt with.

To compare the results to spin-contaminated spin orbital CC, an identical approach to spin error estimations of Hanrath and Engels-Putzka \cite{Hanrath_2009} was used. Given the projection operator $\hat{\mathcal{P}}_{\text{CSF}}^{S,S_z}$ onto the full CSF basis for the desired spin quantum numbers $S$ and $S_z$, an error estimate $\epsilon$ is given by 
\begin{equation}
\epsilon = \sqrt{1 - \Braket{\Psi_{\text{CC}}^{S,S_z} | \Psi_{\text{CC}}^{S,S_z}}}\,,
\end{equation}
where $\Ket{\Psi_{\text{CC}}^{S,S_z}}$ shall denote the projected CC wavefunction after normalization ($\braket{\Psi_{\text{CC}} | \Psi_{\text{CC}}} = 1$) with 
\begin{equation}
\Ket{\Psi_{\text{CC}}^{S,S_z}} = \hat{\mathcal{P}}_{\text{CSF}}^{S,S_z}\ket{\Psi_{\text{CC}}}\,.
\end{equation}
To compare the influence of spin completeness, a spin-adapted and spin-incomplete (SASI) CC was conducted
in a completely identical fashion to the corresponding SASC-CC calculation, where all $\hat{E}$ operators including spectating indices were neglected. This leads to a cluster operator $\hat{T}$, which spans the same spatial space while leaving the spin space incomplete.  

The results of test calculations for the high spin states of the Boron atom in the 6-31G basis set are summarized in Table \ref{BoronResults.tab}. 

All SASC-CC and SASI-CC calculations lead to a spin projection error $\epsilon$ of zero (within the double floating point precision limit). Therefore, both the SASC-CC and the SASI-CC wavefunction always represent a true eigenfunction of the $\hat{S}^2$ operator. In direct comparison, spin orbital CC leads to spin projection errors between $10^{-2}$ to $10^{-7}$ for the doublet state and $10^{-4}$ to $10^{-7}$ for the quartet state. The spin projection error ultimately decreases to zero for the full cluster operator (the FCI result). In case of the hextet state, all participating CSFs are purely composed of $\alpha$ electrons such that the results for spin orbital, SASI- and SASC-CC are completely identical.

SASC-CC correlation energies show small differences to spin orbital correlation energies in the order of $10^{-4}$ to $10^{-10}$\,a.u. in the doublet case and $10^{-5}$ to $10^{-12}$\,a.u. in the quartet case. For a full cluster operator (the FCI limit), the SASC-CC correlation energy is identical to the spin orbital CC correlation energy. In all but the CCSDTQ calculations for the doublet and the quartet state, SASC-CC leads to a greater amount of recovered correlation energy compared to spin orbital CC. The exception for CCSDTQ may be explainable by the non-variationality of the CC-ansatz. The SASC-CC wavefunction should (despite a smaller amount of recovered correlation energy) still be superior to the spin orbital wavefunction as clearly indicated by the spin projection error. Please note that the investigated example (atomic Boron in 6-31G) is very small and represents a minimal testing case for the proposed scheme. Therefore, only small spin contamination effects are recognizable. Overall, we expect absolute spin contamination effects and spin errors to increase with increasing system size. Relative errors w.r.t. the correlation energy however, are expected to remain similiar. While the effect of spin adaption on the amount of recovered correlation energy seems to be minor, spin adaption is expected to be more important for molecular properties -- spin-dependent properties in particular. Recently, this issue has been adressed by Datta and Gauss\cite{Datta2019} for the prediction of hyperfine coupling tensors.

Comparing SASC-CC to SASI-CC correlation energies, a significantly large bias of roughly $4\cdot 10^{-4}$\,a.u. (~1.04\%) 

\onecolumngrid
\begin{center}
\begin{table}[h]
\caption{\label{BoronResults.tab}Correlation energies as well as spin projection errors for (i) spin orbital, (ii) spin-adapted and spin-incomplete (SASI) and (iii) spin-adapted and spin-complete (SASC) CC for the Boron atom in the 6-31G [10s,4p]/(3s,2p) basis set. The ROHF-MOs were converged to a density and energy threshold of $10^{-14}$\,a.u. and CC residual mean squares were converged to $10^{-14}$.}
\begin{tabular}{c|cc|cc|cc}
\textbf{Truncation} & \multicolumn{2}{c|}{\textbf{Spin Orbital CC}} & \multicolumn{2}{c|}{\textbf{SASI-CC}} & \multicolumn{2}{c}{\textbf{SASC-CC}} \\
& $E_{\text{corr}}$ [a.u.] & Spin Error $\epsilon$ &  $E_{\text{corr}}$ [a.u.] & Spin Error $\epsilon$ &  $E_{\text{corr}}$ [a.u.] & Spin Error $\epsilon$ \\ \hline
\multicolumn{7}{c}{$^2$P State ($S = S_z = \frac{1}{2}$) $E_{\text{ROHF}} = -24.519\,348\,011\,198\,5$\,a.u.} \\ \hline
S      & $-$0.000\,136\,326\,135\,3 & 1.09e-02 & $+$0.000\,003\,481\,765\,9 & 5.79e-42 & $-$0.000\,354\,917\,438\,0 & 1.74e-18 \\
SD     & $-$0.043\,007\,929\,406\,6 & 1.00e-03 & $-$0.042\,560\,129\,702\,6 & 3.49e-18 & $-$0.043\,011\,099\,401\,8 & 3.48e-18 \\
SDT    & $-$0.043\,542\,073\,861\,8 & 2.96e-05 & $-$0.043\,088\,321\,409\,8 & 3.97e-19 & $-$0.043\,542\,154\,149\,0 & 5.14e-18 \\ 
SDTQ   & $-$0.043\,543\,752\,049\,9 & 8.22e-07 & $-$0.043\,089\,893\,581\,2 & 3.52e-18 & $-$0.043\,543\,751\,825\,6 & 5.40e-18 \\
SDTQ5  & $-$0.043\,543\,757\,474\,4 & 5.78e-16 & $-$0.043\,089\,899\,014\,8 & 5.32e-19 & $-$0.043\,543\,757\,474\,4 & 3.73e-18 \\ \hline
\multicolumn{7}{c}{$^4$P State ($S = S_z = \frac{3}{2}$) $E_{\text{ROHF}} = -24.442\,277\,339\,965\,4$\,a.u.} \\ \hline
S      & $-$0.000\,005\,630\,917\,5 & 5.39e-04 & $+$0.000\,000\,010\,963\,1 & 1.90e-24 & $-$0.000\,039\,736\,326\,1 & 2.17e-19 \\
SD     & $-$0.006\,325\,166\,426\,4 & 8.73e-05 & $-$0.006\,278\,370\,580\,7 & 1.31e-19 & $-$0.006\,325\,487\,910\,9 & 1.96e-19 \\
SDT    & $-$0.006\,333\,018\,560\,2 & 5.92e-06 & $-$0.006\,285\,473\,968\,9 & 3.33e-19 & $-$0.006\,333\,024\,838\,2 & 3.59e-19 \\
SDTQ   & $-$0.006\,332\,986\,673\,2 & 1.15e-07 & $-$0.006\,285\,438\,483\,6 & 1.66e-19 & $-$0.006\,332\,986\,666\,7 & 1.93e-19 \\
SDTQ5  & $-$0.006\,332\,986\,717\,6 & 3.01e-18 & $-$0.006\,285\,438\,520\,5 & 1.74e-19 & $-$0.006\,332\,986\,717\,6 & 2.50e-19 \\ \hline
\multicolumn{7}{c}{$^6$S State ($S = S_z = \frac{5}{2}$) $E_{\text{ROHF}} = -17.554\,641\,076\,098\,1$\,a.u.} \\ \hline
S    & \phantom{$-$}0.000\,000\,000\,000\,0 & 0.00 & \phantom{$-$}0.000\,000\,000\,000\,0 & 0.00 & \phantom{$-$}0.000\,000\,000\,000\,0 & 0.00 \\
SD   & $-$0.006\,003\,148\,033\,4 & 0.00 & $-$0.006\,003\,148\,033\,4 & 0.00 & $-$0.006\,003\,148\,033\,4 & 0.00 \\
SDT  & $-$0.006\,096\,372\,778\,5 & 0.00 & $-$0.006\,096\,372\,778\,5 & 0.00 & $-$0.006\,096\,372\,778\,5 & 0.00 \\
SDTQ & $-$0.006\,093\,894\,161\,4 & 0.00 & $-$0.006\,093\,894\,161\,4 & 0.00 & $-$0.006\,093\,894\,161\,4 & 0.00 \\
\end{tabular} 
\end{table}
\end{center}
\twocolumngrid

in the doublet case and roughly $5\cdot 10^{-5}$\,a.u. (~0.75\%) in the quartet case for the spin-incomplete cluster operator is found through all truncation levels. Even for a cluster operator incorporating up to 5-fold spatial substitutions, the results do not reach the FCI limit. Clearly, spin adaption without spin completeness may lead to undesirable errors in the correlation energy, which may even be inferior to spin-contaminated CC. 

\section{Conclusion}
\label{Conclusion}

A rigorous scheme to generate linearly independent and spin-complete spatial substitution operators $\hat{E}$ of arbitrary substitution rank and for arbitrary high spin references with spin quantum number $S$ was developed. The proposed scheme utilizes L\"owdin's projection operator method \cite{Lowdin1955, Lowdin1964} of spin eigenfunction generation to ensure the spin completeness of the generated operators. In direct comparison to other open-shell CC methods (e.g. COSCC \cite{Datta2008, Datta2013} or orthogonally spin-adapted CC \cite{Li1993, Li1994, Li1995}), the generated cluster operator in this work is composed of spatial substitutions composed of sole $\hat{E}$ operators only.

The proposed scheme uses a four step procedure (described in section \ref{Implementation}) to arrive at a final set of linearly independent and spin-complete spatial substitution operators. These steps include
\begin{itemize}
\item[(i)] the generation of all $\hat{E}$ operator prototypes of a specific rank, which lead to different spatial functions when applied to the reference CSF (c.f. \ref{PrototypeGeneration}),
\item[(iii)] the generation of primitive spin functions ($S_z$ eigenfunctions), which lead to linearly independent CSFs when used in L\"owdin's projection operator method (c.f. \ref{IndexPermutation}),
\item[(ii)] the mapping of spin particle to spatial orbital permutations employing a topological approach (c.f. \ref{SpinSpatialMapping}) and
\item[(iv)] the application of a canonical index ordering for orbital spaces $\mathbb{O}$, $\mathbb{V}$ and $\mathbb{A}$ (c.f. \ref{IndexSymmetry}).
\end{itemize}

All of the four steps combined act as a black box procedure, which when given a spatial substitution rank $m$ returns a set of linearly independent and spin-complete $\hat{E}$ operators. These are valid for arbitrary high spin references and produce non-orthogonal CSFs when applied to the reference CSF. In our ongoing studies (to be published), we also developed an orthogonalization routine for the generated spatial substitution operators. Our current results point to no noticable difference between orthogonal and non-orthogonal operator sets within the double precision limit.

The proposed scheme was checked for completeness and linear independency for up to 10-fold substitutions and multiplicities of up to $2S+1 = 11$ (c.f. section \ref{Application}) by explicitly evaluating matrix ranks of CSF representations in determinant bases. The number of the generated operators were confirmed by the Weyl-Robinson dimension formula (see e.g. \cite{Pauncz1979}).

A proof of concept CC implementation using the generated operators was developed and successfully applied to the high spin states of the Boron atom (c.f. \ref{TestCalculation}). The spin-adapted and spin-complete CC leads to spin projection errors of zero and to small differences in correlation energies when compared to spin-contaminated spin orbital CC. A comparison to spin-adapted but spin-incomplete CC showed a persistent error of 0.4\,mH to 0.05\,mH in the correlation energy.   

\section{Data Availability}
The data that support the findings of this study are either available within the article or available from the corresponding author upon reasonable request.

\bibliographystyle{LitStyle}

\end{document}